\documentclass[sigconf]{acmart} 
\AtBeginDocument{%
  }

\begin{document}

\title{“I can do whatever I put my mind to!”}
\subtitle{How prioritizing belonging in the design of technology-based programs can support foster-involved youth with self-efficacy, self-expression, and personal exploration.}

\author{Ila Krishna Kumar}
\email{ilak@mit.edu}
\orcid{0000-0002-3482-6615}
\authornotemark[1]
\affiliation{%
  \institution{Massachusetts Institute of Technology}
  \city{Cambridge}
  \state{Massachusetts}
  \country{USA}
}

\author{Karishma Chadha}
\email{kchadha@mit.edu}
\orcid{0009-0002-1901-6259}
\affiliation{%
  \institution{Massachusetts Institute of Technology}
  \city{Cambridge}
  \state{Massachusetts}
  \country{USA}
}

\authornote{Both authors contributed equally to this research.}

\renewcommand{\shortauthors}{Kumar \& Chadha}

\begin{abstract}
In this paper, we examine how intentionally designing for belonging impacts outcomes in a technology program serving a population of youth who have not had safe and supported experiences with technology---youth impacted by foster care. We reflect on the design of an internship program which engaged two foster-involved youth in creative self-expression and co-design with technology. We first outline how we designed the program to foster belonging, then analyze intern and facilitator reflections to understand how the program impacted the interns' experience. We highlight how the internship design affected youth's inclusion and overall engagement; self-expression; exploration; and self-efficacy. We conclude with a discussion of design recommendations for technologists and service providers aiming to engage youth in expressive and co-design technology experiences, particularly youth impacted by foster care. 

\end{abstract}


  


\maketitle

\section{Introduction}
\begin{quote}“I can do whatever I put my mind to!” 
\end{quote}

\begin{quote}“I have come a long way with controlling my emotions and not giving up so fast.”
\end{quote}


\begin{quote}“I am always welcome here.”
\end{quote}

These were reflections shared by young people at the end of a technology-based summer internship for young people impacted by foster care. In this program, youth navigated unfamiliar environments and experiences while exploring technology to express their complex identities, emotions, lived experiences, and aspirations for their futures. Their reflections were particularly notable because, earlier in the program, they had shared that coming to the internship each day felt like “entering a different world,” and that throughout the program, they had each navigated significant moments of self-doubt, internalized frustrations, and complex emotions connected to their identities, relationships, personal goals, and life experiences. In this paper, we surface how prioritizing belonging in the design of technology-based programs can impact young people's self-expression, exploration of their interests and goals, and self-efficacy.

Scholars have highlighted that creative experiences with technology can open up engaging and impactful opportunities for young people to express themselves, connect with others, and explore their identities and interests \cite{papertChildrensMachineRethinking1993, resnickLifelongKindergartenCultivating2018}. In particular, researchers have explored how supporting open-ended, hands-on, and self-directed experiences for young people to tinker, explore, design, and create with technology can foster their sense of agency, problem solving, and critical thinking, \cite{papertChildrensMachineRethinking1993, resnickLifelongKindergartenCultivating2018} which are especially important given the increasing ubiquity of technology in our society \cite{faverioTeensInternetDevice2025}. However, not all young people have access to safe and supportive opportunities to explore and create with technology.

We see cultivating belonging as crucial to engaging youth in freely and safely creating and exploring with technology. Research has identified belonging as a fundamental human need \cite{baumeisterNeedBelongDesire1995}, and a key ingredient in fostering well-being, motivation, and self-determination \cite{baumeisterNeedBelongDesire1995, ryanSelfdeterminationTheoryFacilitation2000}. However, little is understood or agreed upon about how to cultivate belonging \cite{allen_belonging_2021}. HCI research has primary examined how sense of belonging affects community participation \cite{yoon_its_2025,nisi_connected_2023,modi_finding_2025, bustos_ode_2024}, rather than exploring design elements required to support belonging. The little research on how youth technology programs impact aspects of belonging highlight that these programs often do not explicitly consider the design elements required to foster inclusion (a core component of belonging) \cite{oguine_inclusion_2025}. In our work, we outline how we intentionally designed a technology-based program to foster "moments of belonging" \cite{wiseDesignBelongingHow2022} for youth, and examine how this impacted their experience. 

Specifically, we analyze the impact of an internship program designed to support youth in the foster care system with expressing themselves and co-designing with creative technology. We focus on what it means to cultivate belonging for foster-involved youth as this is a community that is particularly in need of supportive and safe environments to express themselves and feel a sense of agency in their lives. First, we outline the ways in which we intentionally designed for belonging in the internship's core infrastructure and activities. We then identify how the program impacted youth's general comfort and sense of inclusion in an unfamiliar environment as well as their self-expression, exploration, and sense of self-efficacy. We end with recommendations for technologists and service providers who are looking to foster belonging while engaging youth in self-expressive and co-design experiences with technology, particularly for young people impacted by foster care.

\section{Related Works}
We describe needs of foster-involved youth and the potential value of supporting this population with technology-based creative expression and co-design. We also outline methods that have been used towards these goals in the past, and how we extend these approaches in the internship's design. 

\subsection{The Needs of Foster-Involved Youth}

In the United States, over 500,000 youth enter the child welfare system every year, because the government identifies them as being at risk of abuse or neglect in their homes \cite{StatelevelDataUnderstanding}. Approximately 20\% of these young people are deemed to be at significant risk of future harm, are removed from their homes, and placed in temporary, state-supervised homes, a situation known as foster care \cite{StatelevelDataUnderstanding}. Although the foster care system is intended to ensure youth’s safety and well-being, research shows that young people are actually at risk of additional abuse or neglect by those responsible for their care in the system. \cite{clarkShownLoveBrokenness2024, biehalMaltreatmentFosterCare2014, garciaMitigatingBarriersImplementing2019, kirknerSexualAbuseDisclosure2024, semanchinjonesSchoolDefinitelyFailed2018} The chronic stressors youth experience before entering and while in foster care can lead to developmental trauma, negatively impacting young people’s cognitive abilities; interpersonal skills; ability to build a healthy sense of self and plan for the future; and abilities to process, regulate, and express their emotions  \cite{blausteinTreatingTraumaticStress2019}. As a result, research shows that over 50\% of these young people transition into adulthood with at least one mental illness \cite{marke.courtneyMentalHealthSubstance2025} and 30-50\% without the social support they need to cope with challenges and reach their goals \cite{okpychMemoCalYOUTHDifferences2018}. 

Experts who support the healing and well-being of youth impacted by developmental trauma advocate for interventions that build foundational emotion identification and modulation skills, healthy social connections, and a positive and cohesive sense of self \cite{blausteinTreatingTraumaticStress2019}. For adolescents, they encourage programs and tools to center young people’s privacy, autonomy, peer support, and connection \cite{blausteinTreatingTraumaticStress2019}. Psychologists who study holistic well-being also speak to the importance of having experiences of exploration, growth, and control over one’s environment \cite{kaufmanTranscendNewScience2020}, which may be particularly important for young adults in foster care who have had little freedom and decision-making power \cite{geenenTomorrowAnotherProblem2007, powersPerspectivesYouthFoster2018}.

Unfortunately, youth in foster care currently have few opportunities to safely explore their environment, interests, and identities in ways that prepare them for independence. Youth in care are often not given a say in where they live and what services they receive, and are not given developmentally necessary opportunities to practice independently making choices, addressing challenges, and setting goals \cite{geenenTomorrowAnotherProblem2007, powersPerspectivesYouthFoster2018}. While research suggests that some foster-involved youth are already using digital tools to form and express their identities, self-soothe in response to challenges, and act independently \cite{sageSystematicReviewInternet2022}, they appear to largely do this without support and sometimes covertly, in response to caregiver restrictions. This contributes to high rates of exposure to exploitation, harassment and other harms in online settings \cite{sageSystematicReviewInternet2022}. Scholars highlight a need for more research to explore how to support youth in leveraging digital technologies to positively build their sense of self, ability to navigate their environment, and confidence in their ability to set their own goals and make life decisions \cite{gomezConnectivityExploringInfluence2025, sageSystematicReviewInternet2022}. 

\subsection{The Value of Fostering Creative Self-Expression with Technology}

Constructionist \cite{papertChildrensMachineRethinking1993} and creative learning \cite{resnickLifelongKindergartenCultivating2018} frameworks advocate for supporting young people in creatively expressing themselves with technology, particularly through hands-on experiences that support them in designing and creating personally meaningful artifacts. In particular, these frameworks emphasize the importance of designing creative technology experiences to enable multiple pathways for expression, guided by young people's personal identities and interests \cite{resnickLifelongKindergartenCultivating2018}. 

Little research has examined how to support foster-involved youth in using technology to authentically express themselves in self-directed ways. Related studies explore how foster-involved youth communicate their experiences and seek support online \cite{ammariFindingUnderstandingSupport2025, fowlerFosteringCommunicationCharacterizing2022, freedHelpseekingCopingStrategies2025}, and outline how online spaces can expose trauma-impacted young people to additional potentially traumatic experiences \cite{freedHelpseekingCopingStrategies2025, oguine_inclusion_2025}. A few studies have examined how foster-involved youth can be supported in expressing their emotions and experiences using technology, either through writing \cite{kumarCultivatingSupportiveSphere2025} or visual expression \cite{kumarConnectingComicsDesign2025}. However, these interventions only engage youth in self-expression through limited mediums.

Research outside of technology highlights the power of poetry \cite{clarkYouthAreNot2023}, drama therapy \cite{nsonwuFosterCareChronicles2015}, photography \cite{capous-desyllasUsingPhotovoiceMethodology2019}, and other arts-based activities \cite{coholicHelpfulnessHolisticArts2009} for supporting the self-reflection, self-expression, and self-efficacy of foster-involved youth. This connects to expressive arts therapy frameworks, which describe how creative, nonverbal methods of expression can be very helpful for processing and expressing emotional experiences that developmental trauma makes difficult to name \cite{malchiodiCreativeInterventionsTraumatized2015}. In this work, we design a technology-based program to support foster-involved youth with creatively expressing themselves in self-guided ways. 

\subsection{The Impact of Engaging Youth in Technology Co-Design Experiences}

Being able to explore, grow, and develop a sense of purpose is important for reaching holistic well-being \cite{kaufmanTranscendNewScience2020}. For youth who have faced significant challenges, supporting them in seeing themselves as experts who have the ability to challenge inequities and manifest a more just future for others who have faced similar challenges may be particularly impactful \cite{makercastroCriticalConsciousnessWellbeing2022}. Youth in foster care may especially benefit from these decision-making opportunities, as they are often given few opportunities to decide what happens to them or even give feedback on services they receive \cite{salazarAuthenticallyEngagingYouth2021}. 

Centering youth voice in the design of technology not only supports the well-being of youth co-designers, it also leads to technologies that more effectively take youths’ needs, barriers, culture, and concerns into account and results in more impactful, engaging and accessible interventions. \cite{willmottParticipatoryDesignApplication2023, folkLessonsLearnedCoDesigning2025, meheraliAdolescentVoicesAction2025, sockolowAtriskAdolescentsExperts2017, stiles-shieldsIncreasingDigitalMental2023} Experts emphasize that effective technology co-design experiences ensure that youth have decision-making power, feel psychologically safe, can choose how and when to participate, understand their rights and responsibilities, and directly benefit from the process. \cite{ezimoraReflectionsFosterYouth2025, folkLessonsLearnedCoDesigning2025, porcheResearchersPerspectivesDigital2022, salazarAuthenticallyEngagingYouth2021, thellTraumaSupportApp2025, wessellsPromotingVoiceAgency2021} 

The few prior works on co-designing technologies with foster-involved youth have focused on co-creating apps and interfaces for youth to get peer support, therapeutic services, and other mental health resources \cite{ezimoraReflectionsFosterYouth2025, kumarConnectingComicsDesign2025, kumarCultivatingSupportiveSphere2025}. While these studies incorporate youth perspectives in the design process, youth were largely providing input rather than making their own design decisions. In our program, we aimed to support young people in completing a self-directed technology design project that would help other foster-involved youth with fully expressing themselves. 

\subsection{Designing for Belonging in HCI}

While belonging has been defined in many ways \cite{allenBelongingReviewConceptual2021}, we adopt a description articulated by designer Susie Wise as encompassing the feelings of being accepted, invited to participate, connected with others, and authentically oneself \cite{wiseDesignBelongingHow2022}. Technology research has explored how belonging connects to academic engagement in computing environments \cite{blaney_examining_2017,veilleux_role_2012,kaneria_belonging_2024, runa_role_2025,von_briesen_interventions_2025} and provides examples of custom digital interfaces \cite{nisi_connected_2023}, online communities \cite{yoon_its_2025, modi_finding_2025}, and art programs that foster a sense of belonging  \cite{bustos_ode_2024}. Little research has explored how designers can cultivate belonging for youth in open-ended experiences with technology, especially outside of environments they are already in. The most related research, examining existing inclusion practices of youth makerspaces, highlights a lack of programs that utilize intentional design choices to foster inclusion and the need for specific design recommendations on how to foster inclusion in technology-based youth programs \cite{oguine_inclusion_2025}. Our work addresses this gap by sharing specific design strategies for fostering belonging in technology-based experiences and examining their impacts on youth. 


\section {Program Context}

\subsection{Internship Program Conceptualization}
The internship grew out of partnerships with three organizations that serve foster-involved youth in the United States. Two of these organizations serve youth in Pennsylvania, offering statewide “Child Preparation” (Child Prep) services \cite{pennsylvaniastatewideadoptionandpermanencynetworkswanpermanencytoolkitSWANPermanencyToolkit2026} which aim to help youth process their lived experiences through  (1) finding personal records and information about their life history and (2) expressing aspects of their identity, interests, experiences, and personal stories through creative activities. The third organization offers a paid summer internship program for foster-involved youth in Massachusetts. This program connects young adults with local organizations and businesses to build personal savings, professional skills, and work experience. With these partners, we saw an opportunity to host a paid youth internship centered on creative expression with technology and the co-design of new creative and expressive activities to support other foster-involved youth.

\subsection{Facilitation Team}
There were two core facilitators of the internship program, both of whom also subsequently co-authored this paper. Neither have lived experience with foster care. Both facilitators are graduate student researchers in the research group that hosted the internship. One facilitator is a queer South Asian American person whose research centers on supporting youth who have experienced developmental trauma with exploring, designing, and creatively expressing themselves with technology. They also work as an active Court Appointed Special Advocate (CASA), focusing on helping young adults work towards self-determined goals and be seen, heard, and valued by their care team and community. They were motivated to engage in this project because of a strong belief in the importance of creating safe spaces for youth who have experienced significant early life challenges to express themselves and experience agency. The other lead facilitator of this paper is a South Asian American graduate student whose research focuses on designing welcoming experiences for young people to create and express themselves with technology. She designs digital tools and creative workshops that aim to foster identity-affirming experiences for young people, particularly from communities that have often been excluded from creative opportunities with technology. She additionally examines the ways existing and emerging technologies can better support belonging for young people of all backgrounds.

\section{Program Design}

We hosted a 6-week summer internship program for foster-involved youth. We describe our core design considerations for the internship program, particularly as they relate to cultivating a sense of belonging. In the following sections, we detail aspects of the program design organized by the specific "moments of belonging" \cite{wiseDesignBelongingHow2022}---moments during community participation that can impact one's sense of belonging---that they connect to.

\subsection{Invitations}

Through how we presented the internship and engaged with potential interns, we tried to signal that intern candidates were valued and welcome in technology settings, and could be themselves and follow their unique passions in this space. 

\subsubsection{Job description} We began by crafting a job description (Appendix \ref{job-description}), which our partner organization shared with youth in their network. We intentionally highlighted our focus on supporting the interns in feeling comfortable being themselves in the space, experiencing agency, and pursuing their personal interests. We intended for this to come across in the outlined internship goals (e.g. "You experience the joy of creating things that you care about."), the expectations outlined for interns (e.g. "Pay attention to how you're feeling."), and the dress code (e.g. "You should feel free to wear whatever makes you comfortable").

\subsubsection{Interviews} We conducted a set of interviews with 7 interested youth at our research lab. In the interviews, we emphasized that each candidate belonged in contexts like the internship program. We did this by starting the interview with an overview of the internship, only asking questions about what the candidate was excited about after ensuring that they had a strong grasp of the goals and core activities of the program. We also verbally emphasized to all candidates that they deserved to be in spaces like the internship program and hosting research lab, highlighting specific strengths we noticed in their responses in real-time. 


\subsection{Entering}

We intentionally set up the internship environment to help the interns feel invited into the space and program. Much of this was prepared before the interns' first day, to provide them with immediate, tangible evidence that they were welcome, cared for, and encouraged to exercise their agency. 

\subsubsection{Personal desk space} 
\begin{figure}[t]
    \centering
    \begin{minipage}[b]{0.28\textwidth}
        \includegraphics[width=\textwidth]{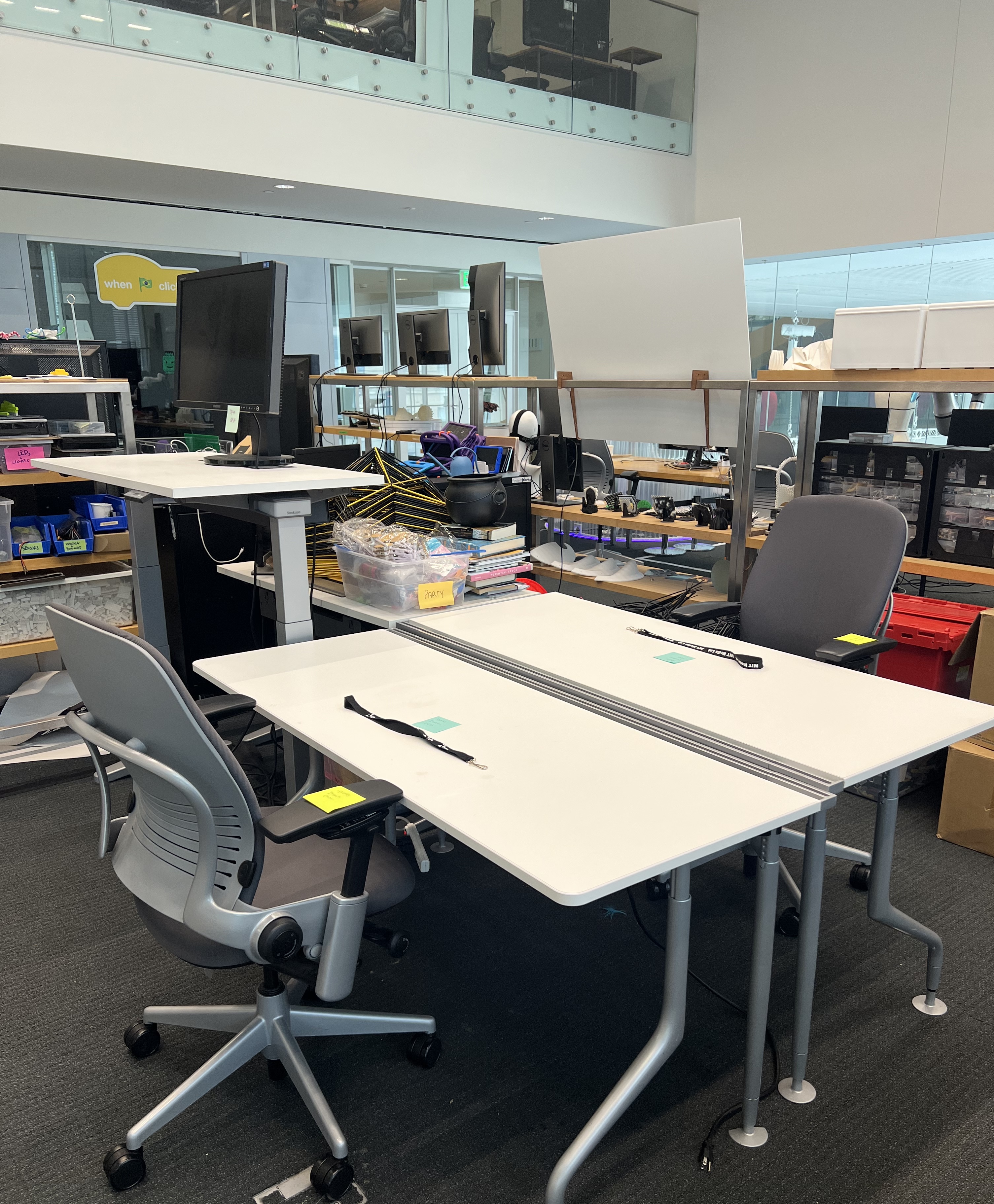}
        \caption{Intern desks, before they started.}
        \Description{An image of two white rectangular desks facing one another, with rolling chairs pushed in. The desks appear to be in an open office space, next to other desks and supplies.}
        \label{fig:desks}
    \end{minipage}
    \hspace{.5cm}
    \begin{minipage}[b]{0.4\textwidth}
        \includegraphics[width=\textwidth]{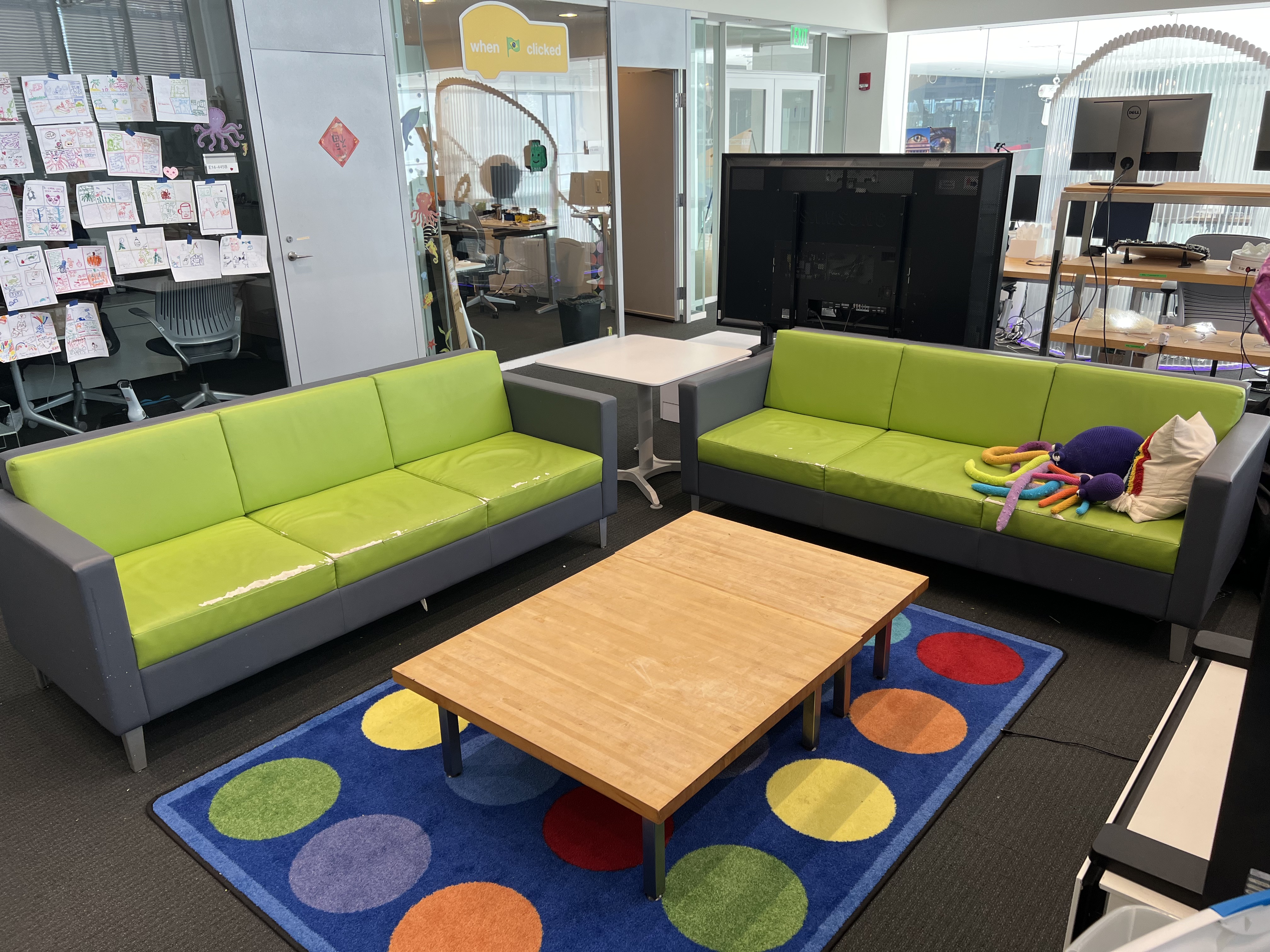}
        \caption{Couches where morning check-ins and end-of-day reflections took place.}
        \Description{An image of green couches perpendicular to one another with a coffee table and colorful rug in the middle.}
        \label{fig:couches}
    \end{minipage}
\end{figure}

\begin{figure}[t]
    \centering
    \begin{minipage}[b]{0.2\textwidth}
        \vspace{0pt}
        \includegraphics[width=\textwidth]{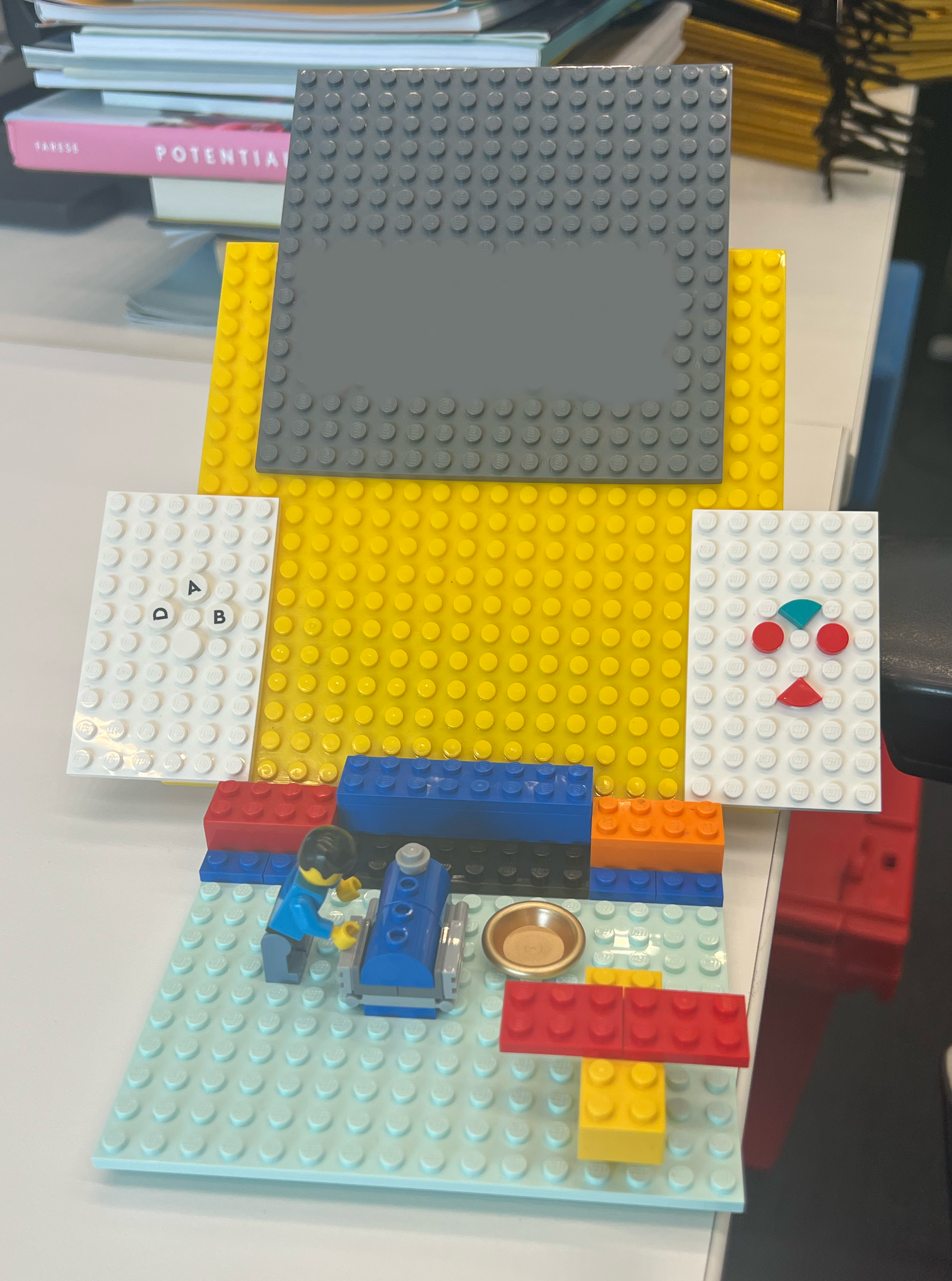}
    \end{minipage}
    \hspace{.5cm}
    \begin{minipage}[b]{0.27\textwidth}
        \vspace{0pt}
        \includegraphics[width=\textwidth]{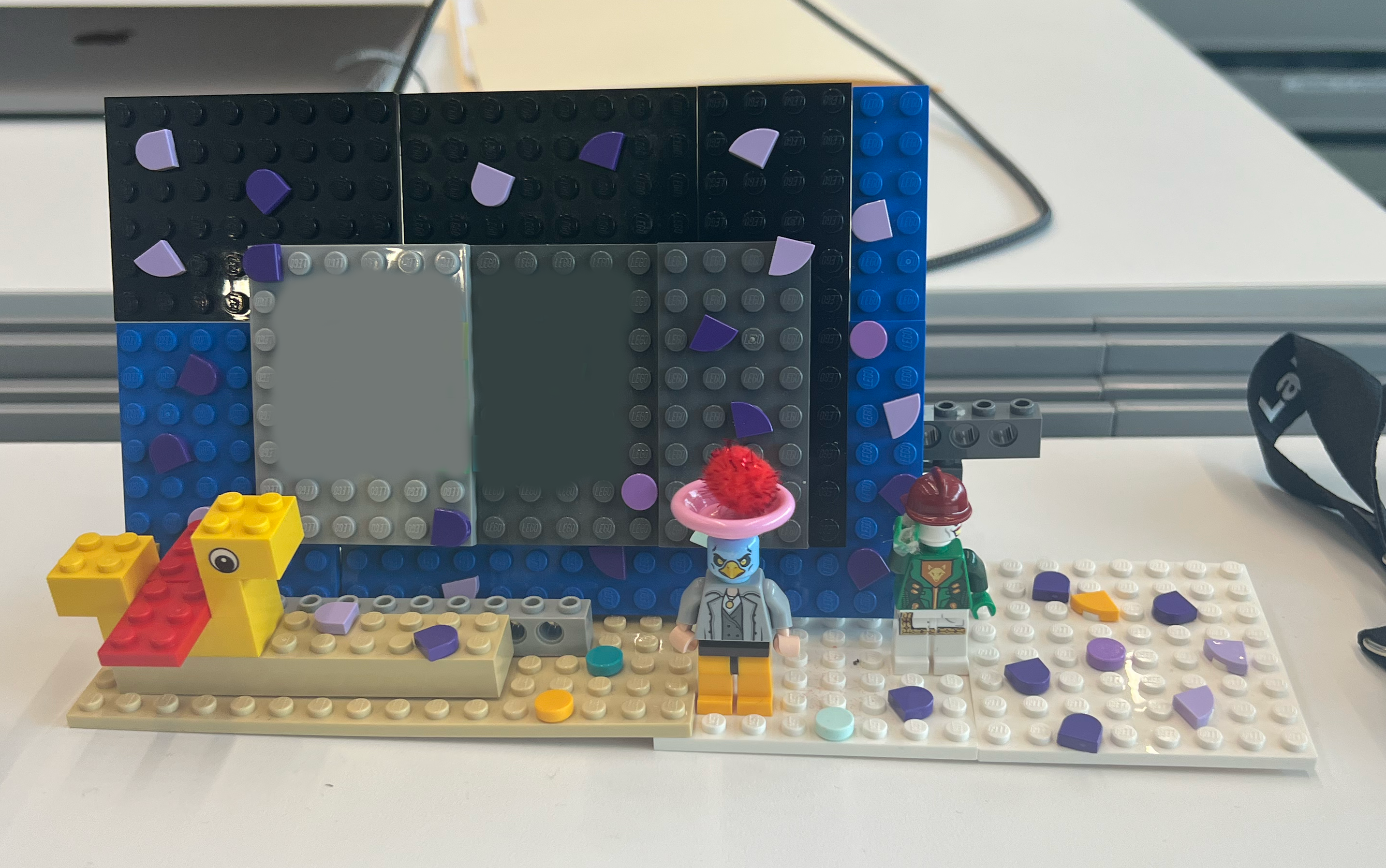}
    \end{minipage}
    \caption{Name plates created by the interns for their desks (with names removed).}
    \Description{Two images of flat LEGO brick boards set up to display an intern's name along with decorations made of LEGO bricks and some other materials, such as a pompom.}
    \label{fig:nameplates}
\end{figure}

Our research group workspace had an open office layout. We set up two dedicated desks (Figure \ref{fig:desks}) for the interns to work at in addition to communal group spaces (e.g. a big table where the group gathers for meetings and couches where we often held end of day reflections, Figure \ref{fig:couches}). On the interns' first day of work, we facilitated an activity around creating name plates that represented aspects of their personalities, to make their desks feel like their own. We did this activity using LEGO bricks to also get the interns engaged in creating and expressing themselves with familiar physical materials (Figure \ref{fig:nameplates}). 

\subsubsection{Unmediated access to food} We felt it was important that interns had food accessible to them at all times. This included arranging for lunch to be provided before their first paychecks came in and providing convenient, unmediated, and free access to snacks, drinks and microwaveable meals. Throughout the internship we asked the interns for their feedback and restocked the selections based on their preferences. 

\subsubsection{Schedule of planned activities} We provided a printed schedule of planned activities for the week, so the interns could mentally prepare for what going to happen and feel a sense of control over their day (Appendix \ref{schedule}). In later weeks, this included open time in which the interns could schedule informal chats with other members of the research group or work on their self-paced design projects. 

\subsubsection{Autonomous access to the space} We worked with our home institution to give the interns official designations, ID cards, and building access so they could freely move in, out, and around the building. Because this process could not be completed until the start of the internship, we printed interim name badges for the interns using the lab's branding. We hoped the IDs (temporary then official) would foster a sense of belonging as members of the lab. A colleague additionally helped us set up a professional photoshoot for the interns so that they would have high quality headshots for their ID cards. We also circulated their headshots in a lab-wide newsletter to formally announce their arrival, to pave the way for them to have safe and positive experiences with others they might meet.

\subsubsection{Welcome lunch} Every week, our research group hosted a weekly group lunch meeting for all members (10-15 people). In these meetings, each person was asked to share about themselves and what they were working on. During the first week of the internship, we held a special welcome lunch for the interns to introduce them to the group and celebrate their arrival. Instead of the usual  prompt, everyone was asked to share something new or out of their comfort zone they'd done recently or a new hobby they were interested in learning. Anticipating that the interns might feel overwhelmed with new internship experiences, we chose this prompt to encourage others to share relatable experiences of being challenged or feeling like a novice. 

\subsection{Participating} We designed the program's main activities to support interns in engaging as their whole selves. This primarily manifested as: facilitating creative technology workshops that aimed to support the interns with expressing aspects of their identities, emotions, and experiences; organizing technology exploration activities guided by intern's interests and goals; and creating space for the interns to regularly reflect on internship activities.  

\subsubsection{Creative technology workshops for self-expression}\label{creative-technology-workshops} 
\begin{figure}[t]
    \centering
    \begin{minipage}[b]{0.3\linewidth}
        \centering
        \includegraphics[width=\linewidth]{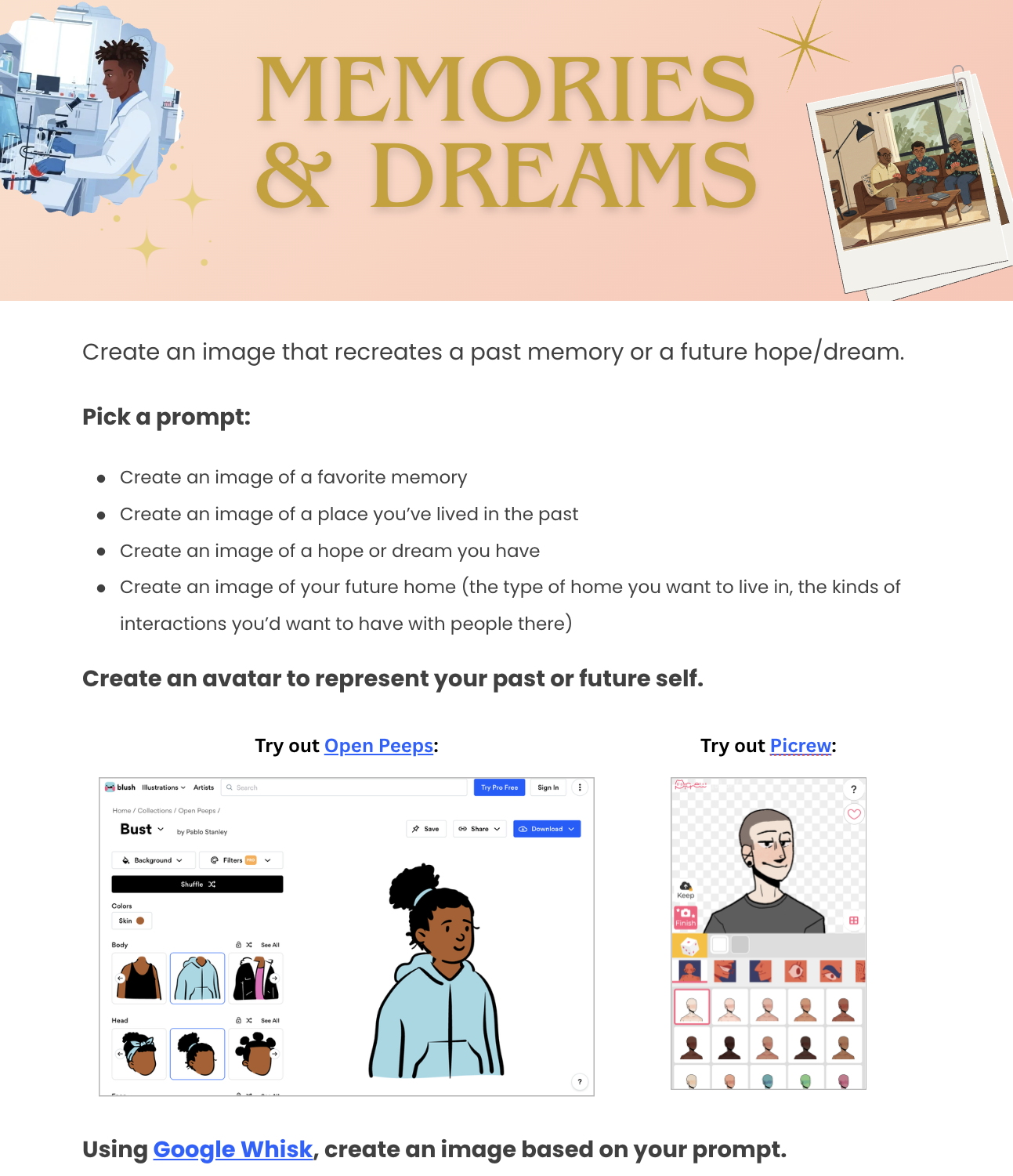}
        \Description{A digital document describing an an activity for using digital graphics tools to create an image connected to past experiences or future goals}
    \end{minipage}
    \begin{minipage}[b]{0.31\linewidth}
        \centering
        \includegraphics[width=\linewidth]{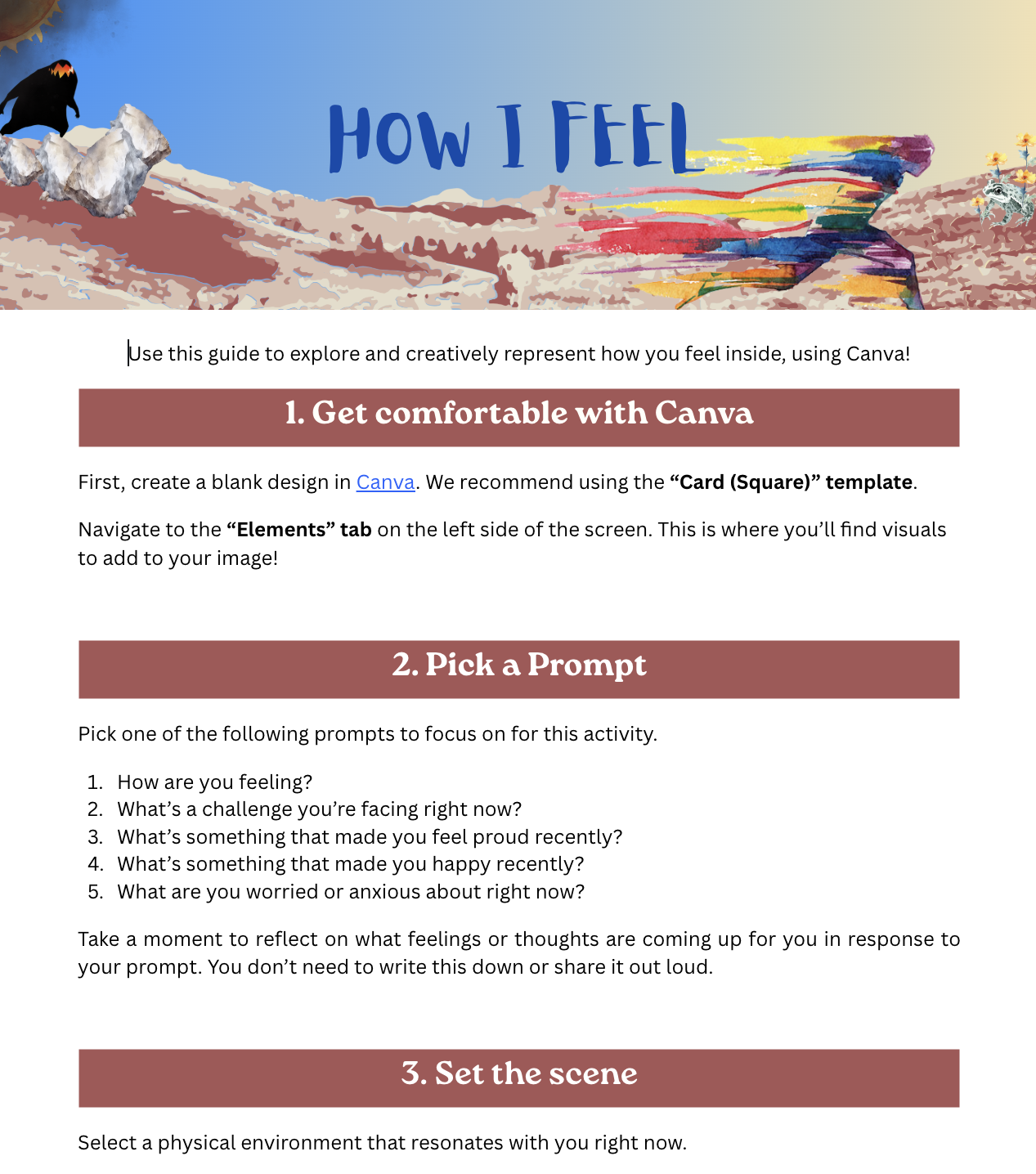}
        \Description{A digital document describing an an activity for using a graphic design tool to visualize emotional experiences}
    \end{minipage}
    \begin{minipage}[b]{0.29\linewidth}
        \centering
        \includegraphics[width=\linewidth]{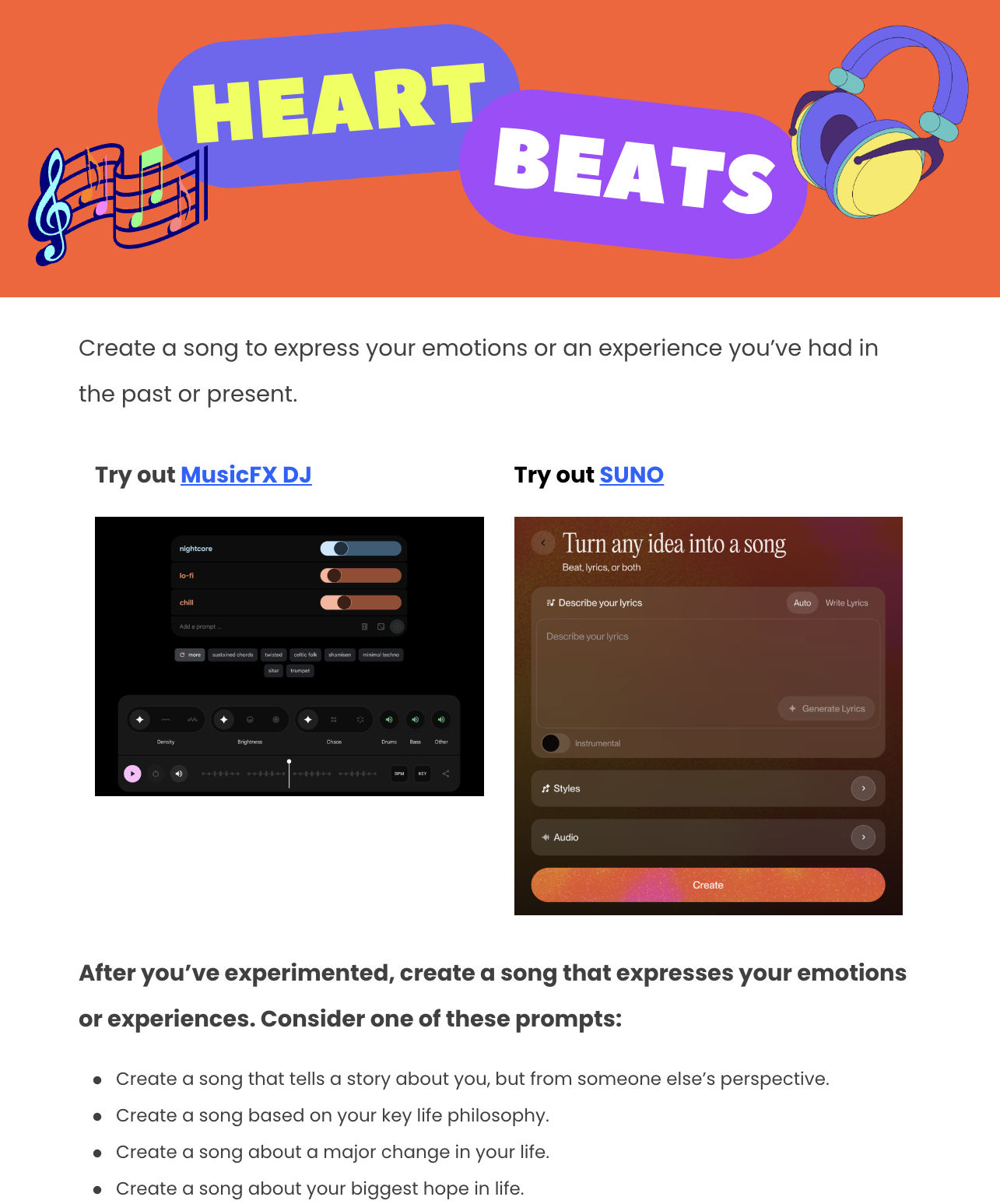}
        \Description{A digital document describing an an activity for using music creation tools to create songs connected to various life experiences}
    \end{minipage}
    \caption{Excerpts from creative digital activity guides for: (1) using Blush and Google Whisk to create a graphic illustrating a memory or dream for the future, (2) visualizing emotions with Canva, and (3) creating music with Google MusicFX DJ and Suno connected to a life experience or emotion}
    \label{fig:digital-creative-activities}
\end{figure}

\begin{figure}
    \centering
    \includegraphics[width=1\linewidth]{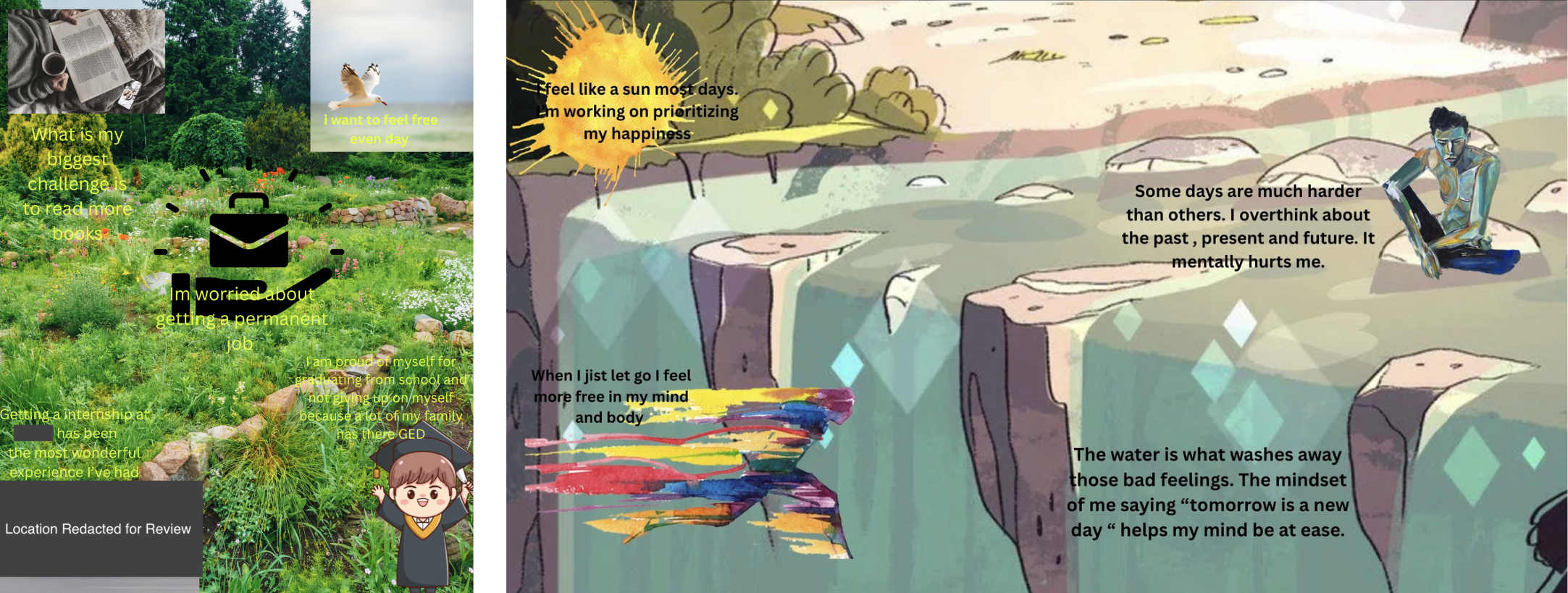}
    \caption{Visualizations created by the interns in response to a "How I Feel" activity (Figure \ref{fig:digital-creative-activities}) expressing personal challenges, emotions, and accomplishments; worries, desires, and hopes for the future; and feelings about the internship.}
    \Description {Two photographs side-by-side. On the left a collage of photographs and graphics overlaid on a background photo of natural greenery. On the right, a collage of digital graphics overlaid on a digital illustration of a waterfall. Text accompanies each of the overlaid graphics on both collages describing what they symbolize to each  intern.}
    \label{fig:how-i-feel-both}
\end{figure}

\begin{figure}
    \centering
    \includegraphics[width=1\linewidth]{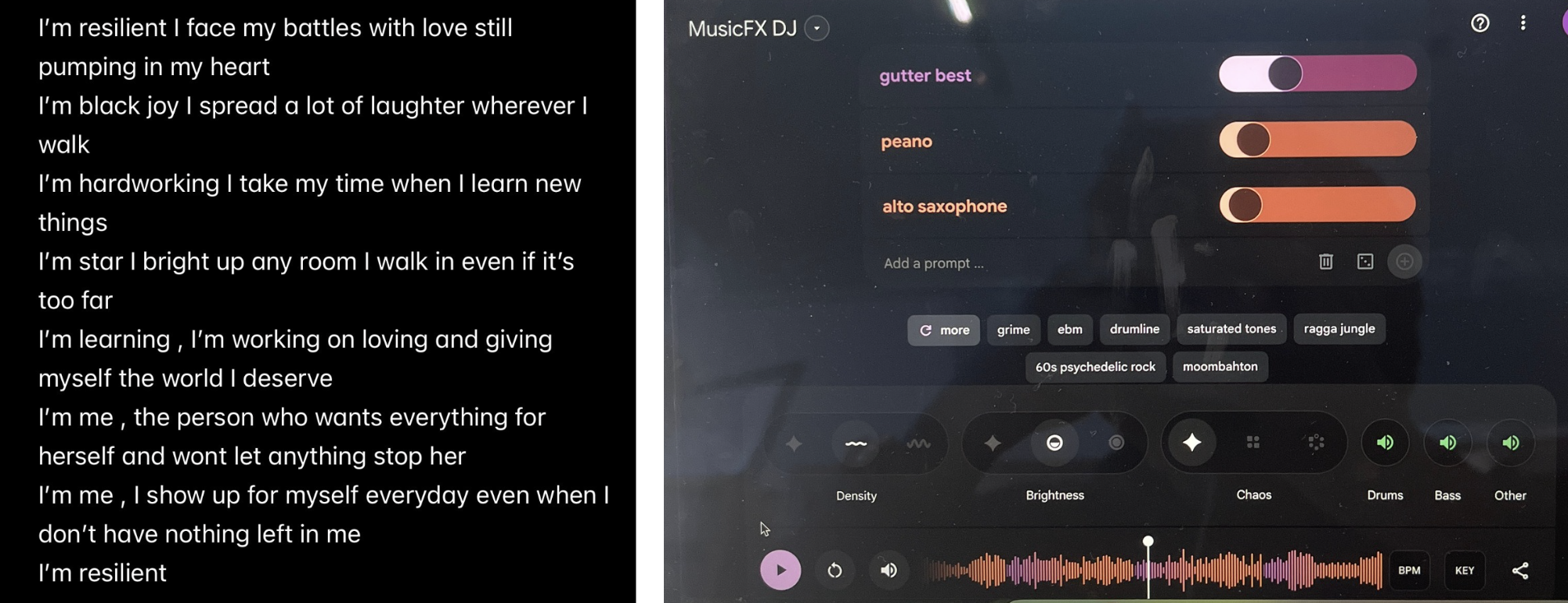}
    \caption{On the left, a poem one intern wrote on the theme of resilience, to use as the lyrics for a song they were generating with Suno. On the right, an intern's in-progress photo of using Music FX DJ to create a song combining music genres and instruments.}
    \Description{Two images side-by-side. On the left a screenshot of a poem with white text on a black background. On the right a photograph of a computer screen with an interface displaying three sliders with different values, with user input prompts as labels "gutter best," "peano," and "alto saxophone." Below the sliders are additional generated prompt suggestions and interface controls labeled "Density," "Brightness," "Chaos," "Drums," "Bass," and "Other" with options to adjust each. At the bottom of the screen is an audio wave form.}
    \label{fig:music-activities}
\end{figure}

\begin{figure}
    \centering
    \includegraphics[width=1\linewidth]{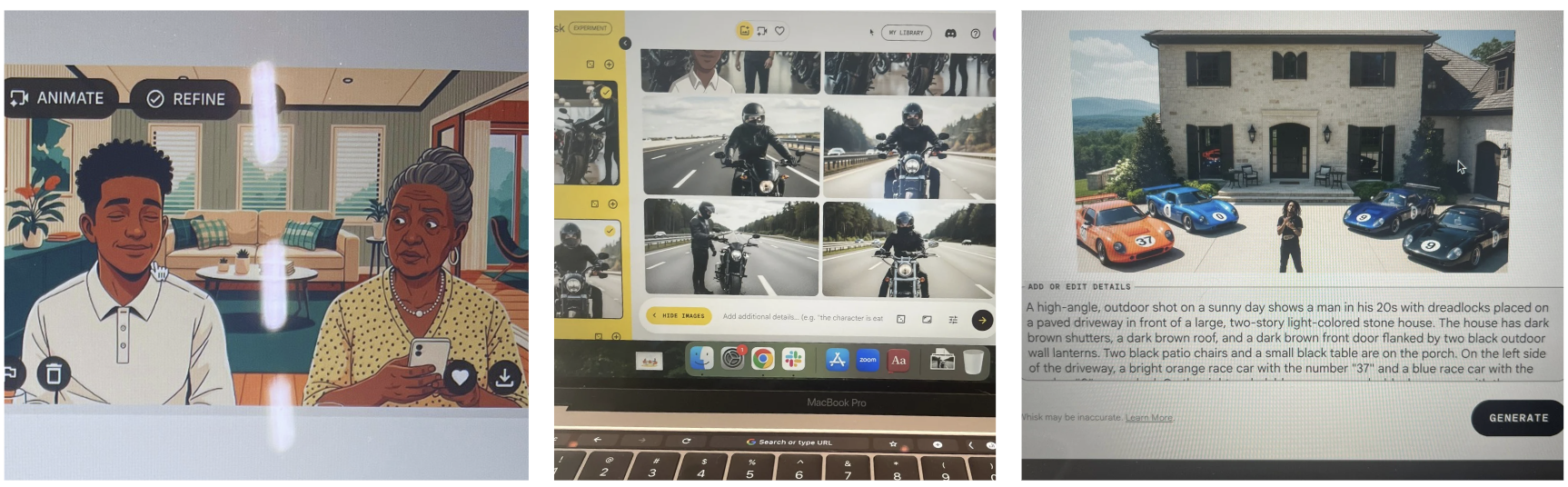}
    \caption{In-progress images of one intern's iterations on visuals created using Google Whisk. The images depict three aspirations for the future.}
    \Description{Three photographs of computer screens displaying the Google Whisk interface of images generated by one intern. The images depict digital illustrations of two people together eating; photograph style images of a person on the road on or with a motorcycle; and a photograph style image of a house with race cars and a person in the foreground.}
    \label{fig:whisk-future}
\end{figure}

We facilitated creative workshops to support interns in expressing themselves with diverse, creative technologies. Creative technologies included Canva \cite{Canva}, a graphic design tool; Blush \cite{BlushIllustrationsEveryone}, a digital avatar creator; generative AI tools for images (Google Whisk \cite{Whisk}) and music (Google MusicFX DJ \cite{MusicFXDJ} and Suno \cite{SunoAIMusic}), and apps developed to support personal and emotional expression (\cite{rusk_honoring_2024, kumarConnectingComicsDesign2025}). With each of these tools, we introduced prompts around themes of self-reflection and self-representation, such as imagining a dream or vision for the future, visualizing your current emotions, and creating music to reflect on past experiences (Figure \ref{fig:digital-creative-activities}). Through these workshops, interns explored complex emotional and personal themes combining different technologies and mediums of expression such as poetry (Figure \ref{fig:music-activities}), visual art (Figures \ref{fig:how-i-feel-both} and \ref{fig:whisk-future}), and music (Figure \ref{fig:music-activities}).

\begin{figure}[t]
    \centering
    \begin{minipage}[b]{0.3\textwidth}
        \vspace{0pt}
        \includegraphics[width=\textwidth]{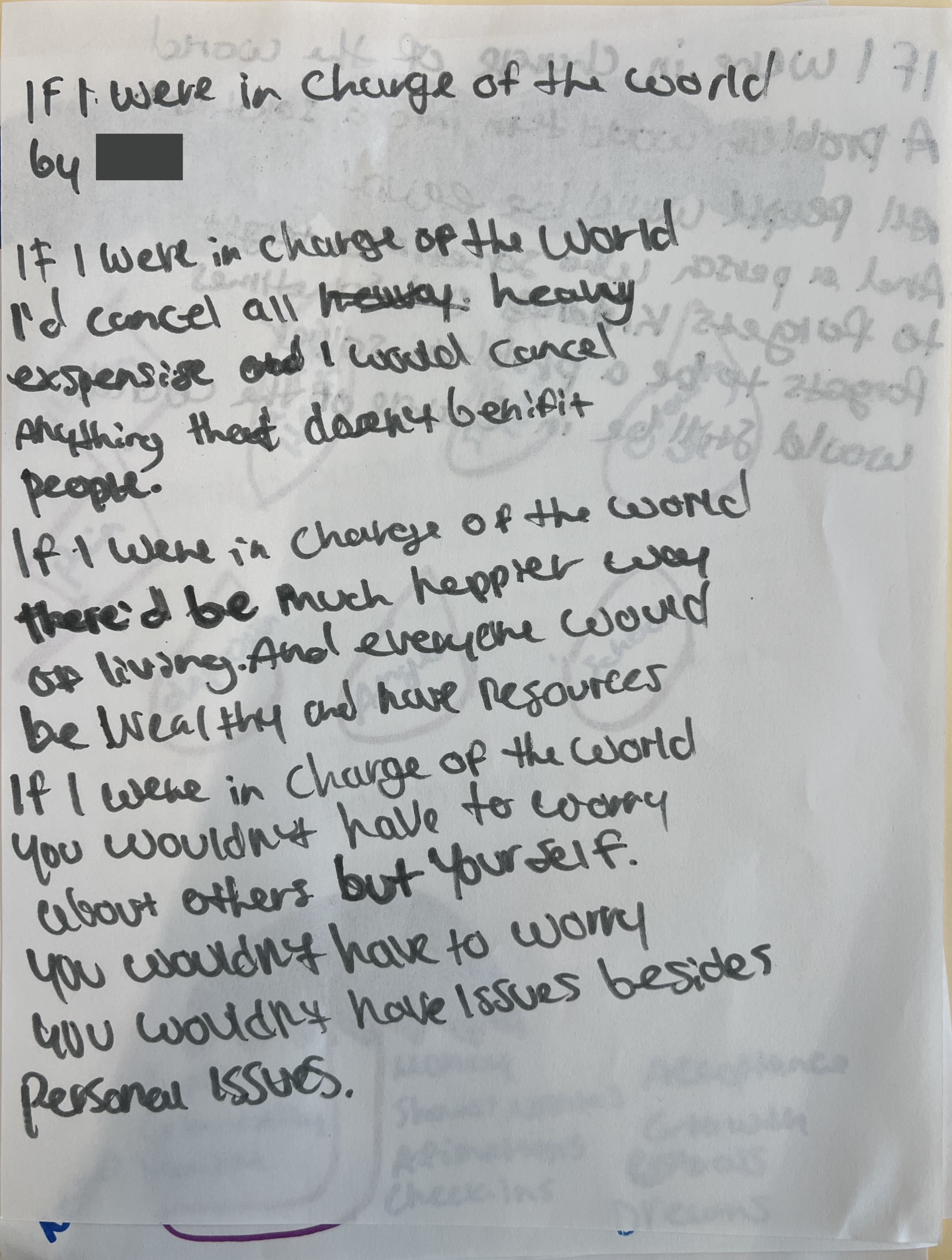}
    \end{minipage}
    \begin{minipage}[b]{0.315\textwidth}
        \vspace{0pt}
        \includegraphics[width=\textwidth]{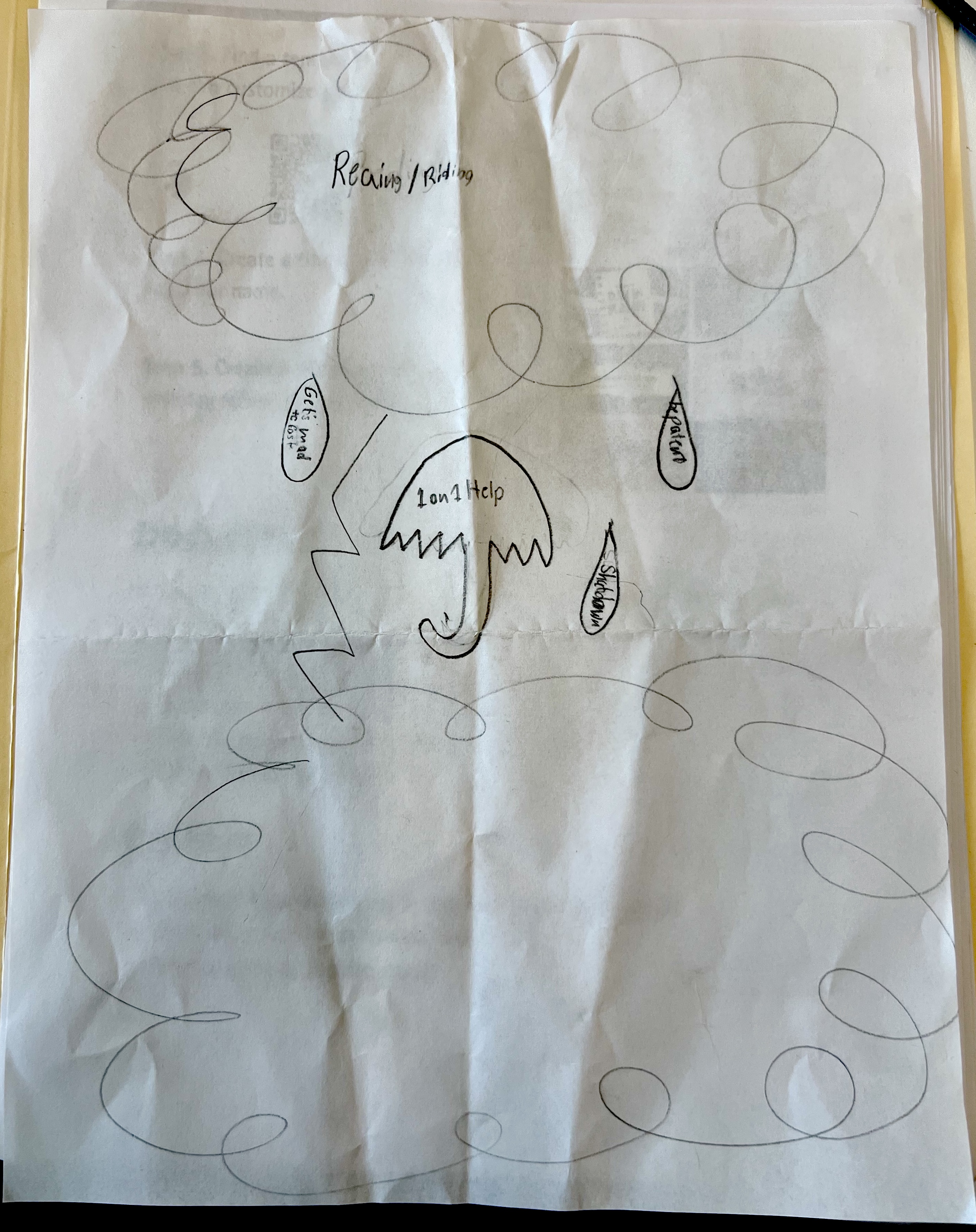}
    \end{minipage}
    \caption{Examples of pieces created by interns in response to existing Child Prep activities. The first piece is a poem using a starter template, called "If I was in Charge of the World." The second piece is a drawing that uses a rainy day metaphor to describe personal obstacles (e.g. reading and writing) and what protects them or helps them cope (e.g. 1 on 1 help).}
    \Description{Two images of white pieces of paper, one with a handwritten poem on it and the other with a drawing of a cloud, raindrops, and an umbrella.}
    \label{fig:child-prep-activities}
\end{figure}

\subsubsection{Daily reflection rituals} Each workday began with a group check-in following the prompt of "How are you feeling today?" to provide space for the interns to share what, if anything, they were bringing into the workspace with them (emotions, challenges they were facing outside of work, goals that were on their mind). We also held reflection conversations after each creative activity, to support interns in processing their experience and understand how we should adjust future activities to meet their needs. Additionally, we planned for interns to keep a daily design journal to individually reflect on activities at the end of each day, noting particular moments of learning or questions they were left with. However, based on early feedback from the interns, we replaced the design journal activity with an end-of-day group reflection. Some common prompts for this discussion were "What are you most proud of yourself for today?" and "What advice would you give to someone else who was trying out the activity we did today for the first time?"

\subsubsection{Emergent explorations} We engaged researchers from [our home institution, redacted for review] in facilitating demos and interactive activities to support the interns in exploring diverse ways that technology can manifest, particularly in combination with creative disciplines. These demos and activities spanned across a variety of technology disciplines including robotics, AI, music, and computational art. In these experiences, we created space for the interns to ask questions about the researchers' career and life journeys, as well as share about their own connected interests and experiences. We also made room for serendipitous experiences that were guided by the interns' emerging interests (surfaced to us through internship activities and reflective conversations). Interest-led activities included one-on-one informal meetings that interns set up with colleagues in our research group, field trips to local museums and a recording studio, and workshops led by colleagues (on topics such as career preparation and meditative movement).

\subsection{Contributing}

We aimed to provide clear avenues for the interns to shape both the design of the internship program as well as broader initiatives to support the well-being of other foster-involved youth.  

\subsubsection{Co-creating community norms} In the first week, we dedicated time to co-creating group expectations with the interns (Appendix \ref{community-norms}). We first held a discussion focused on two questions: “What would make you feel safe and welcomed in the space?” and “What should we all do to make others feel safe and welcomed in the space?” We took notes on a large screen, so interns could make changes to what was documented. Then, we converted discussion notes into draft expectations for the interns, for us as supervisors, and for all of us as a community. On a subsequent day we shared the draft and facilitated a group discussion about what should change, modeling this by first making a suggestion ourselves. We edited the expectations accordingly and reprinted them for the interns to keep on their desks.

\subsubsection{Technology co-design project} Early in the internship, the interns explored some of the non-digital activities currently used in Pennsylvania's "Child Prep" program \cite{pennsylvaniastatewideadoptionandpermanencynetworkswanpermanencytoolkitSWANPermanencyToolkit2026} to build an understanding of this program, which the interns would subsequently be designing new digital activities for \ref{fig:child-prep-activities}. Additionally, we designed the prompts used in the creative technology workshops (described in Section \ref{creative-technology-workshops}) to align with the goals of the Child Prep program. We did this to provide the interns with experiences that could help them craft their own perspectives on how creative digital technologies might support other similar youth. 

\begin{figure}[t]
    \centering
    \begin{minipage}[b]{0.5\linewidth}
        \includegraphics[width=\linewidth]{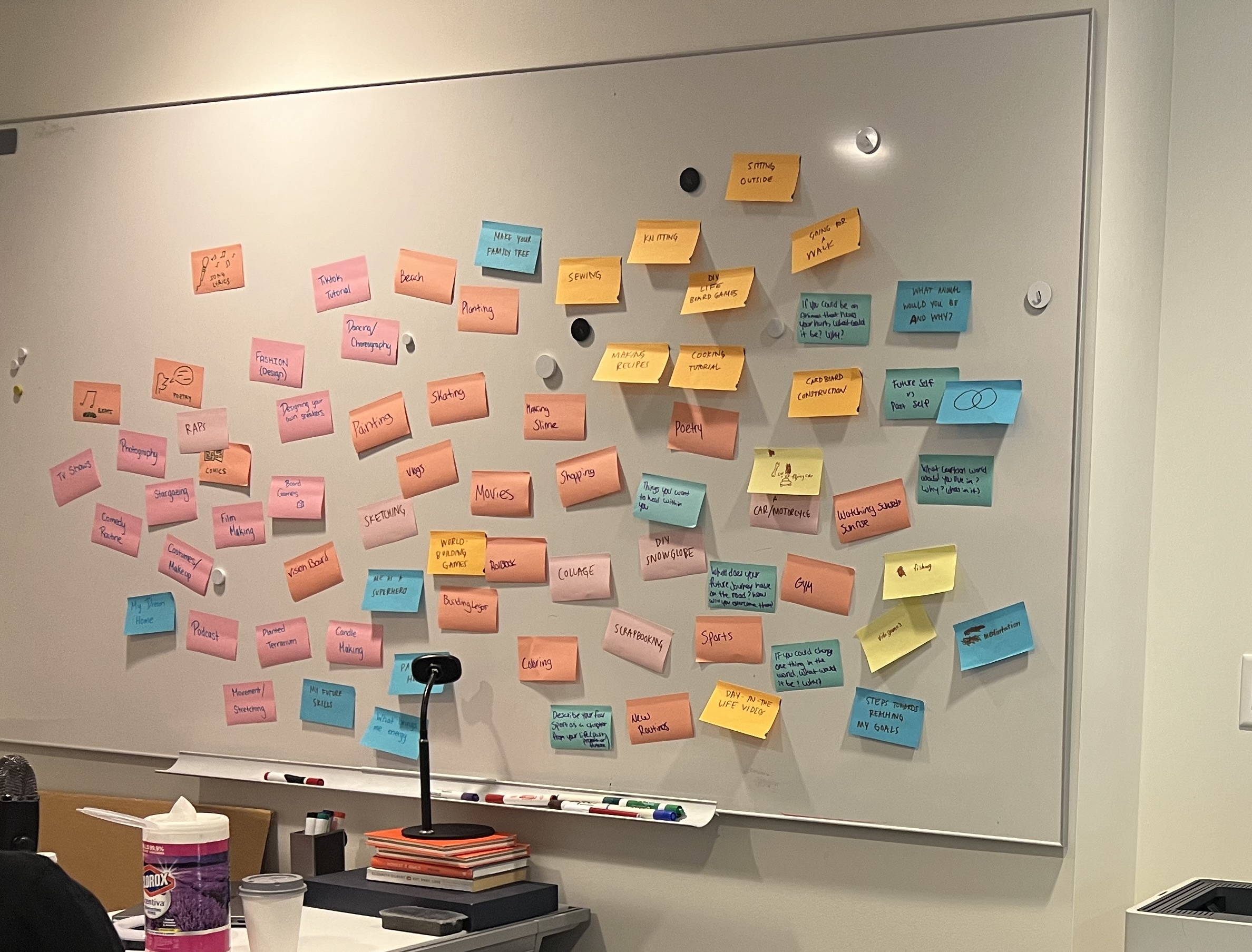}
        \caption{Sticky notes generated during a verbal brainstorm activity.}
        \Description{An image of a white board with many colorful rectangular sticky notes on it.}
        \label{fig:stickynotebrainstorm}
    \end{minipage}
    \hspace{.5cm}
    \begin{minipage}[b]{0.4\linewidth}
        \includegraphics[width=\linewidth]{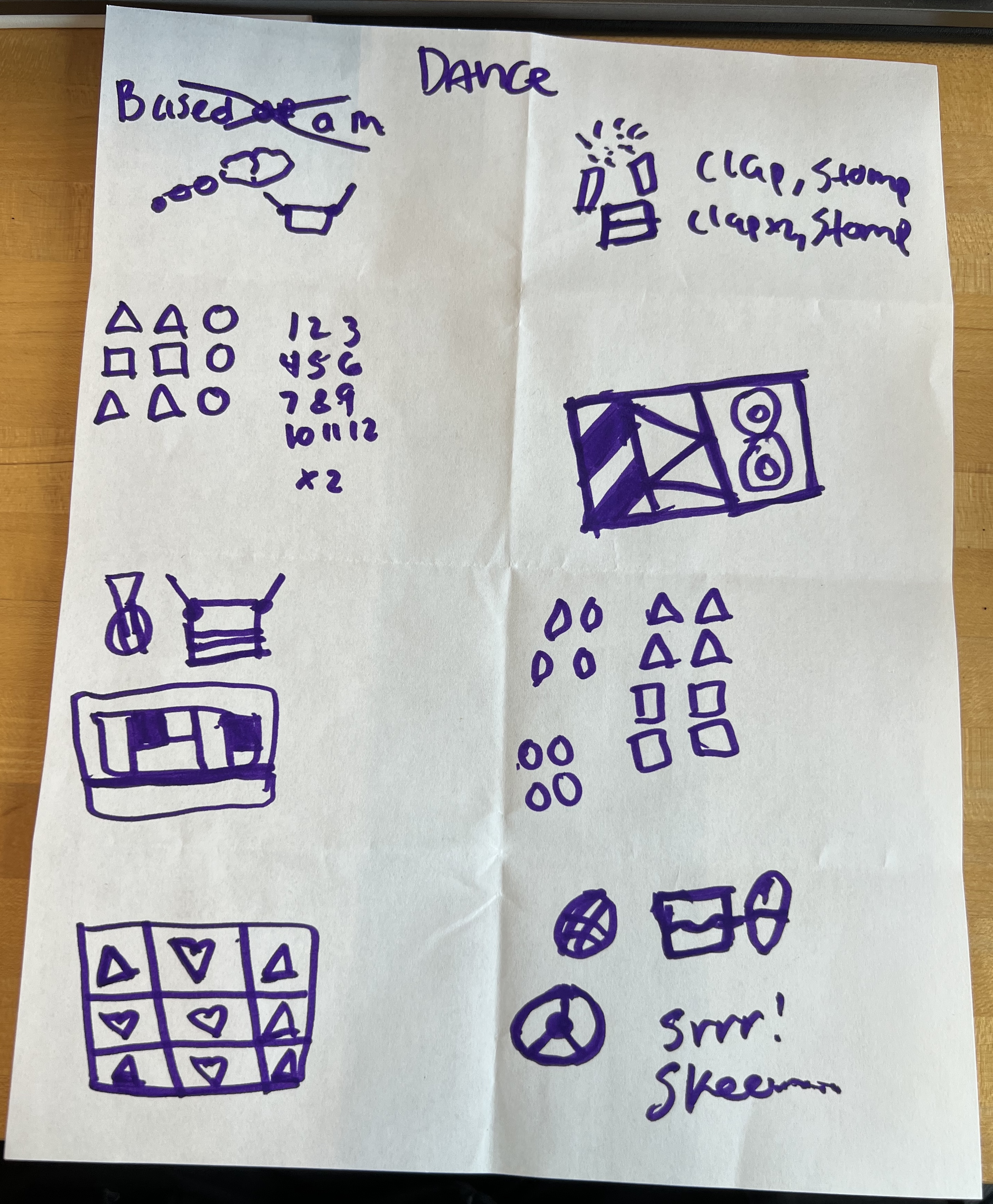}
        \caption{Drawings generated by an interns during a Crazy 8s visual-focused idea generation activity.}
        \Description{An image of a white piece of paper that has 8 sections, each of which containing symbolic drawings. On top is a header that says "Dance".}
        \label{fig:crazy8brainstorm}
    \end{minipage}
    \hspace{.5cm}
    \begin{minipage}[b]{0.6\linewidth}
        \includegraphics[width=\linewidth]{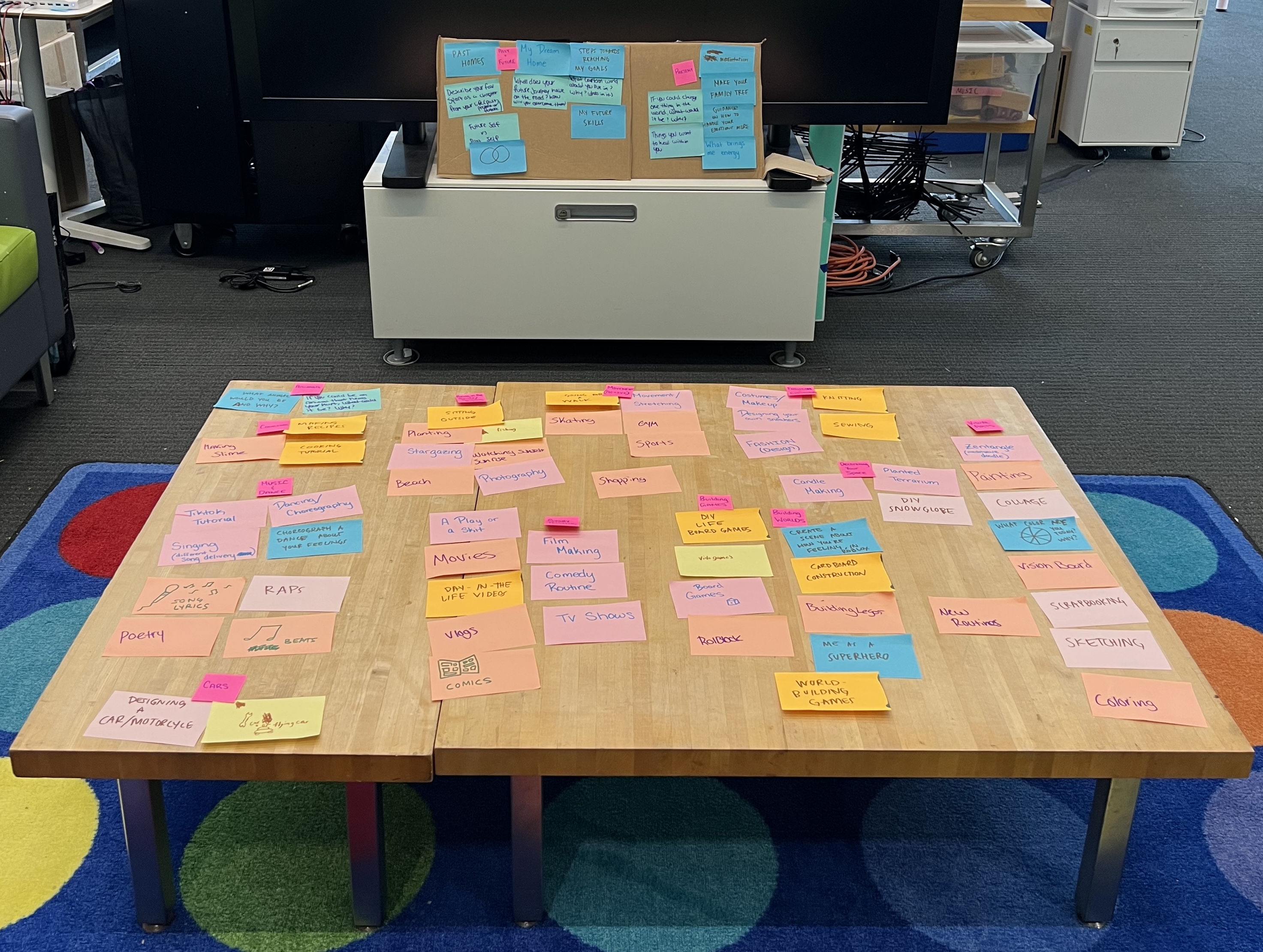}
        \caption{Idea sticky notes categorized into themes for use as inspiration in prototyping new activities.}
        \Description{An image of sticky notes in groups with a smaller sticky note label, spread out on a table and on a piece of cardboard that is propped up on a black TV.}
    \label{fig:ideathemes}
    \end{minipage}
\end{figure}

\begin{figure}[t]
    \centering
    \begin{minipage}[b]{0.3\textwidth}
        \vspace{0pt}
        \includegraphics[width=\textwidth]{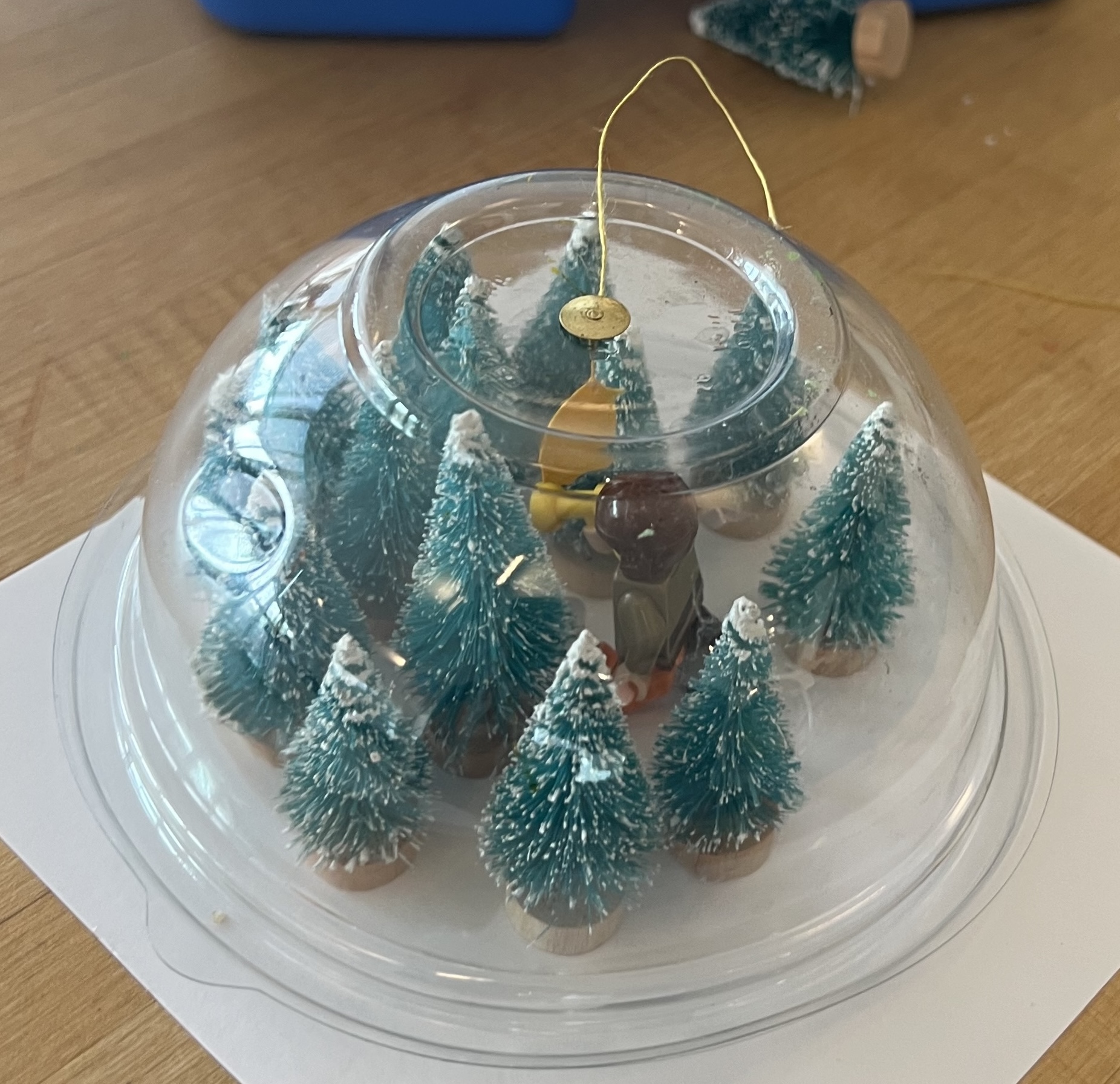}
    \end{minipage}
    \begin{minipage}[b]{0.5\textwidth}
        \vspace{0pt}
        \includegraphics[width=\textwidth]{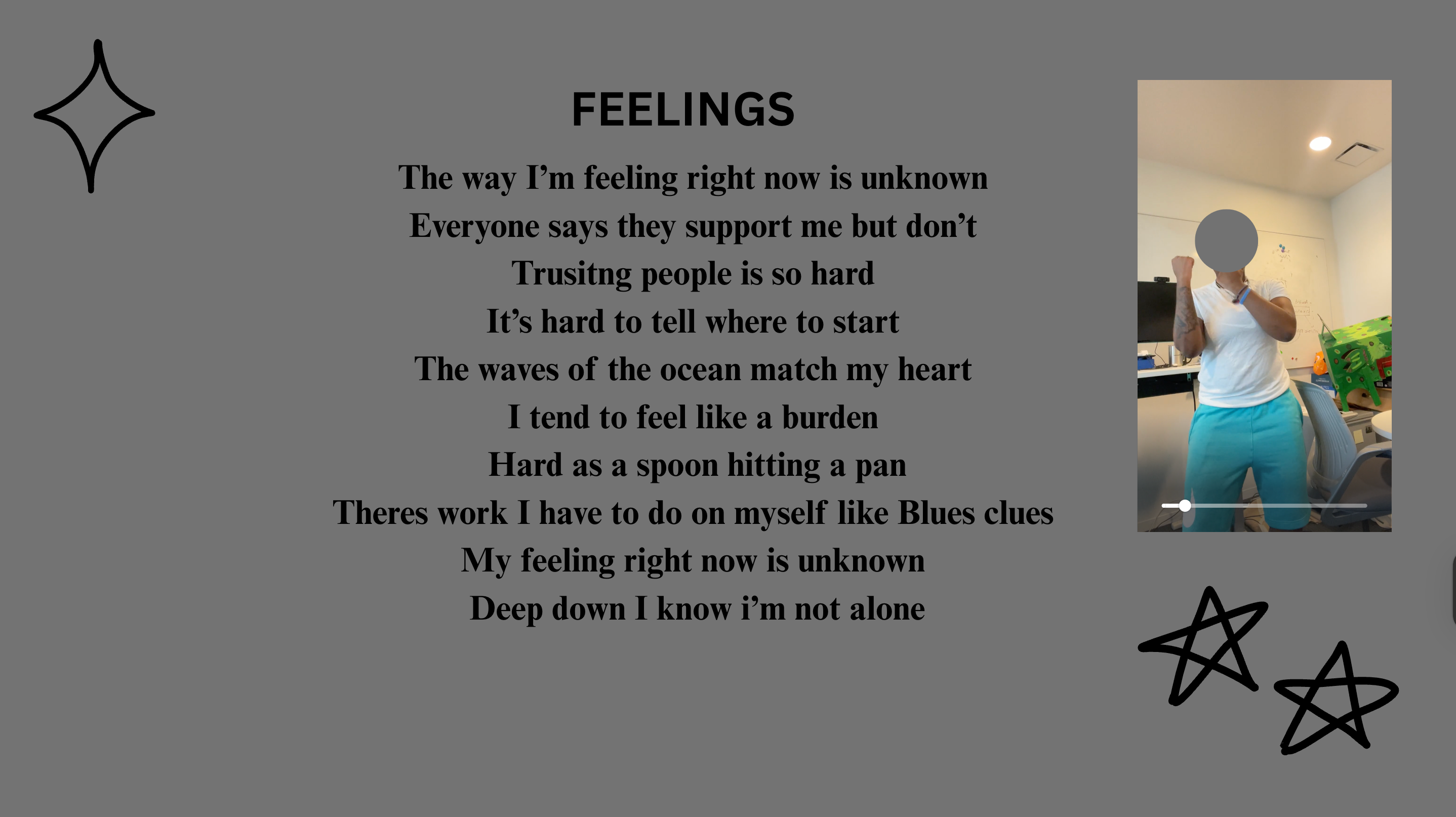}
    \end{minipage}
    \caption{Examples of prototypes created by interns. The first piece was inspired by terrariums, and the intern who created it said it described how they feel sometimes, "cold and lost amongst the trees. The second piece combined poetry and dance to express how they were feeling.}
    \Description{An image of a clear plastic cup on a white piece of paper. Inside, there is a person surrounded by small snow-capped trees. Above them is hanging a yellow circle, which symbolizes the sun. A screenshot of a poem titled "Feelings" on a grey background, with a video in the top right corner of a young person dancing.}
    \label{fig:prototypes}
\end{figure}

In the latter half of the internship, we shifted focus from discrete creative activities towards an open-ended and self-guided process of designing a new creative digital activity for other foster-involved youth. We scaffolded the design process with group brainstorming activities to support the interns in imagining the scope of possibilities, engaging the interns in both verbal (Figure \ref{fig:stickynotebrainstorm}) and visual (Figure \ref{fig:crazy8brainstorm}) forms of brainstorming. We supported the interns in going from divergent to convergent thinking, grouping ideas into themes (Figure \ref{fig:ideathemes}) and digging deeper into concepts they were particularly interested in through prototyping and iteration (for example, Figure \ref{fig:prototypes}). At this stage, we shifted the work day structure to include more unstructured time for the interns to explore and create at their own pace. While we initially planned for the interns to get feedback on their projects with one another or members of their communities outside the internship, due to challenges (discussed in Section \ref{external-life}), the interns ended up primarily sharing their prototypes with us for feedback as they developed their designs.

\subsection{Dissenting \& Repairing}

Through expectation-setting and our facilitation style, we tried to lay the groundwork for a dynamic in which interns felt comfortable sharing critical feedback with us about activities or our actions as facilitators.  

\subsubsection{Intern expectations} From the intern recruitment phase onward, one of the guiding "Expectations" we emphasized was to "Pay attention to how you are feeling." and "If you are comfortable, we are here to listen and figure out how to get you the support you need..." (as can be seen in the job description, in Appendix \ref{job-description} and community norms, in Appendix \ref{community-norms}). With this grounding norm, we tried to foster a safe space for the interns to share their feelings and emphasize that we our priority was for activities to meet their needs. 

\subsubsection{Facilitation style} In conversations with interns, we took great care to encourage critical feedback and receive it with appreciation. In moments where the interns expressed general frustration, we focused on understanding and validating the intern's experience and supporting them in taking care of themselves. We continuously worked on our own emotion regulation and engaged in self-reflection on our facilitation style. When we realized we had made assumptions or mistakes we set aside time to name what we had done and apologize for any harm we caused. In the results, we describe some of these moments and how they impacted the intern's experience. 

\subsection{Exiting}

To end the internship in a way that recognized the accomplishments and growth of each intern, we organized a community celebration and a final reflection activity that left interns with tangible reminders to take with them as we parted ways and they continued working towards their goals. 

\subsubsection{Celebration} 
\begin{figure}
    \centering
    \includegraphics[width=1\linewidth]{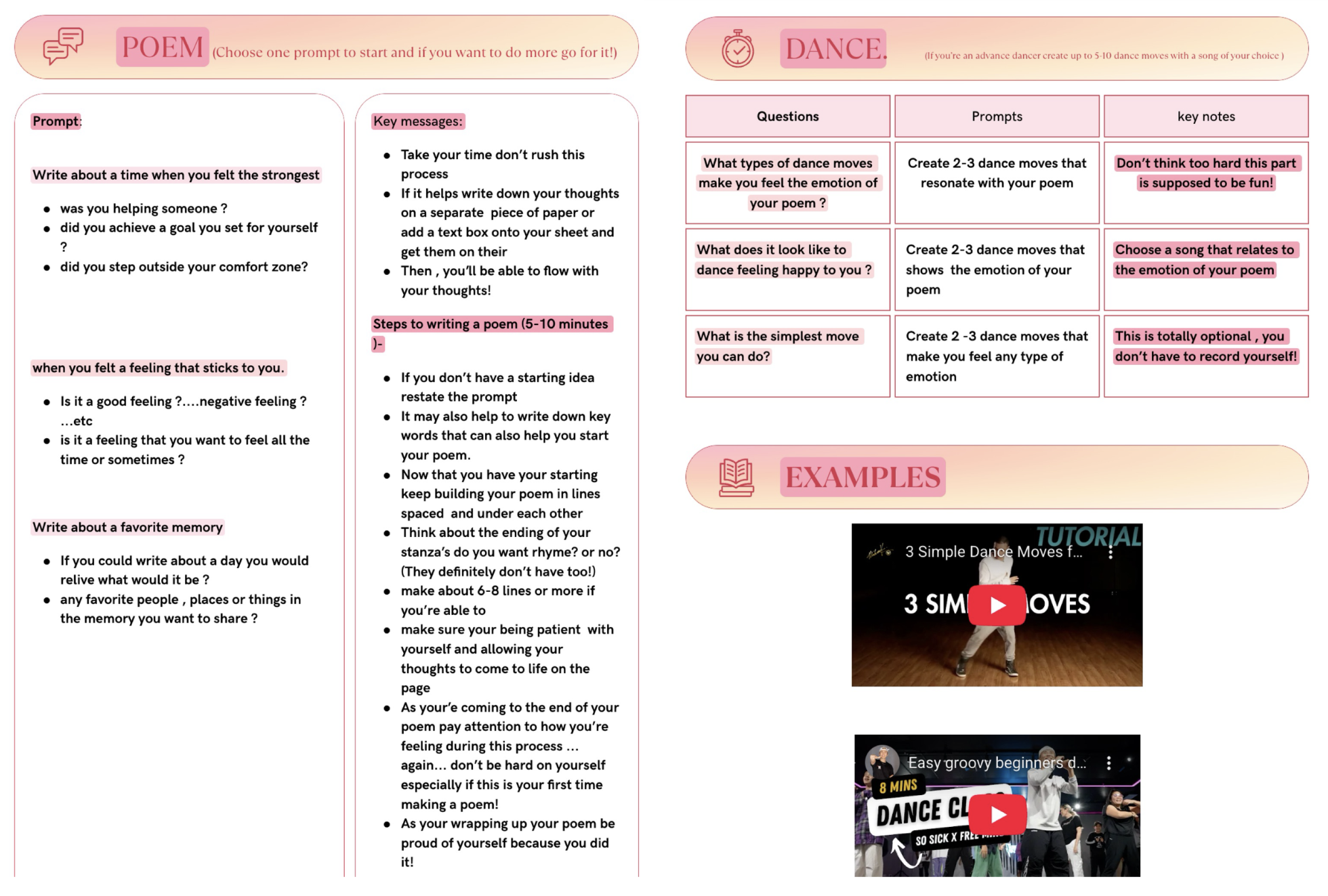}
    \caption{Excerpts from activity designed by one intern exploring poetry, dance, and music}
    \Description{Screenshots of a typed guide about how to create a poem about a feeling or memory and a connected dance. This is followed by stills of two example videos.}
    \label{fig:poem-dance}
\end{figure}

\begin{figure}
    \centering
    \includegraphics[width=1\linewidth]{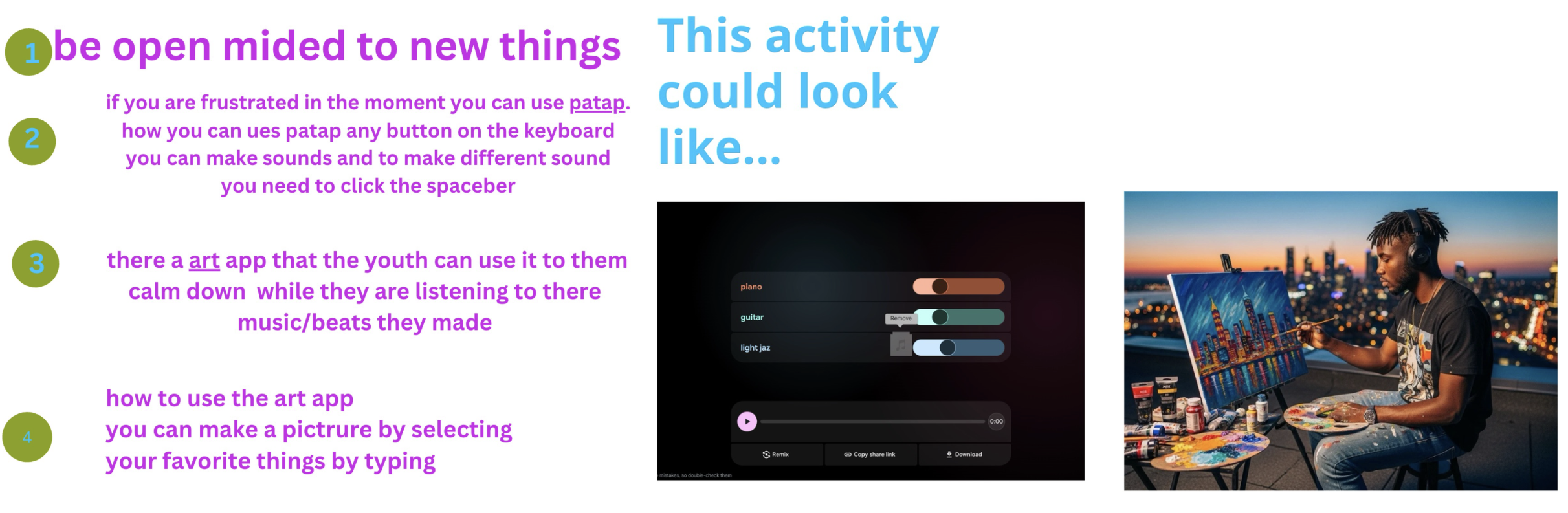}
    \caption{Excerpts from activity designed by one intern exploring art and making beats}
    \Description{Screenshots of a presentation slide deck presenting an activity related to making beats to calm down, and then optionally create a piece of art to go with the creation. This is accompanied by screenshots of a music generation app interface and an example art piece created with a generative art-making tool.}
    \label{fig:art-music}
\end{figure}

We concluded the internship with a celebration attended by members of the research group, colleagues across the lab who had interacted with the interns, and guests the interns wanted to invite from their personal lives. During the celebration, we spoke about the interns' accomplishments and shared our appreciation for them. After this, the interns presented the activities they had designed (Figures \ref{fig:poem-dance} and \ref{fig:art-music}), followed by audience questions about their ideas.

\subsubsection{Final reflection activity} Following celebratory food and desserts, we ended the work day with a final reflection activity, asking the interns to reflect on what they learned, challenges they faced, what they felt they accomplished, what goals and interests they were excited about now, and what takeaways they wanted to remember from the internship. We then facilitated an activity in which the interns designed and created physical button pins based on what they wanted to remember in the future.

\section{Reflection Process \& Analysis Methodology}

We kept a daily log of our observations and reflections. We regularly read each other's notes and discussed our experiences as facilitators to determine how the program activities and our facilitation strategies should be adapted moving forward. We also took photographs during activities and of artifacts created in activities to capture the interns' experiences during the internship. As we discuss earlier, we had reflection conversations with the interns after almost every activity, day, and difficult moment to understand their experience and meet their needs. These reflections were often referenced in facilitator notes. During the final internship reflection activity, we documented direct quotes (contributing to a final quote-based activity, and to aid discussion). After the internship, we, as facilitators, had a retroactive reflection session, individually writing then discussing our takeaways about what we felt had been beneficial and challenging for the interns and for us as facilitators, to process our experiences, improve as facilitators, and organize priorities for future programming. 

After the internship ended, we analyzed de-identified artifacts from internship activities and reflections. Artifacts included de-identified photographs, facilitator reflection logs, digital media created by the interns, early design journal entries, and materials from final project presentations. Our home institution's Institutional Review Board reviewed and approved this retroactive analysis as secondary use research (E-7041). We followed a bottom-up thematic analysis method to identify key themes in the data. Through discussion, we identified four high-level categories within which we grouped the themes: Environmental Factors, Self-Expression, Exploration, and Self-Efficacy.

\section{Results}

We ran the internship program with two youth (ages 18 and 19) in the Massachusetts foster care system. In this section, we share reflections on how the internship design impacted interns' experiences in the program.


\subsection{Comfort, Inclusion, \& Engagement}

The internship and external environments appeared to impact the interns' sense of comfort and inclusion, as well as their affect and engagement in activities. Environmental factors included the workspace setup, the culture of the research group and of the neighborhood in which the host research institution sat, and the life challenges and goals that the interns were navigating outside the internship. 

\subsubsection{Internship environment}

The preemptive actions we took to set up the internship environment appeared to impact how comfortable and included the interns felt in the program. During the first community norm-setting conversation, the interns shared that they already felt safe and welcomed in the internship, in part because we had provided "personal space" that was not crowded or shared, and continuous access to snacks. Over the internship period, we noticed that they used their desks as a place to rest and unwind when they arrived early, during their lunch break, and in gaps between activities. Additionally, although we did not get direct feedback about their perceptions of the ID badges, building access, and how we introduced them across the lab, the interns seemed excited to get their IDs and consistently wore them around their necks and seemed to feel comfortable coming in, out, and walking around the building on their own (even doing so as a way of coping with negative emotions, as described in Section \ref{reflective-conversations}). 

At the same time, the interns shared that "getting used to the environment" was a challenge they worked to overcome during the program. In the final reflection discussion, they elaborated that they learned to get used to "how social [the environment] was, the buildings/architecture, the vibes, [and] the people." They shared that they had to "[get] used to being in a place where everyone likes being there," noting that this environment was different from other environments they had encountered in the past. They also mentioned feeling overwhelmed by the environment, saying "it was overstimulating at first getting used to so many people" and that "talking in groups" felt challenging, especially in the first week of the program. This was likely connected to their experience in the welcome lunch we organized in their first week, as well the fact that our research group had an open office layout, a social culture, and gathered weekly for large group lunch meetings. When referring to the environment being overstimulating and sharing about the challenge of "talking in groups" we believe the interns were referring in part to this social environment being a new experience.

The interns also reflected on how it was challenging "to navigate who I am in this workspace, being true to myself, having an open mind to spread my wings." This may have been due in part to the fact that this was the first time either intern had worked in an office or lab environment. It may also have been connected to an earlier observation the interns made about feeling like traveling from their neighborhood to the internship site felt like "entering a different world," with dramatically different scenery and types of people. Despite these obstacles, the interns appeared to end the program feeling affirmed in their sense of belonging in the space. When asked what they wanted to remember from the internship experience, they said, "I am always welcome here" and "Don’t worry about the noise. I am accepted and I belong here" and elaborated that "here" meant "in places out of [their] comfort zone."

\subsubsection{External environment} \label{external-life}
The interns told us about life challenges, interests, and goals that were frequently on their mind while they were participating in internship activities. This included challenging interpersonal dynamics they were coping with, injuries they were healing from, bureaucratic processes they were navigating to achieve personal goals (such as getting identification documents, passing driving and placement tests, and applying for school financial aid), and the many appointments and calls they were juggling related to these topics. We often learned about these life happenings in morning check-ins and end-of-day reflections, as well as in reflection conversations after activities that brought up emotional reactions. The interns also sometimes shared and reflected on these themes as they engaged in the creative digital activities that invited them to reflect on their feelings, interests, and future aspirations (for example, Figures \ref{fig:how-i-feel-both} and \ref{fig:whisk-future}). We also observed many instances where one intern sharing their goals (e.g. getting a driver's license) inspired the other to reflect on similar goals and interests.

The interns also appeared to be coping with significant amounts of uncertainty and concern about what their life would look like after the internship (including how they would earn money or adjust to new educational programs). These factors caused interns to miss significant portions of the program for meetings and calls, which in turn meant that we had to skip some aspects of the program. For example, near the end of the internship, one intern missed multiple days of work because they were meeting with different support staff to finish paperwork for a vocational program they wanted to start in the following month. Because of the absences and inconsistent attendance, the interns were unable to try out each other's design prototypes, resulting in not being able to give each other feedback and iterate on their designs. Since there were only two interns and they had built a connection with one another, one's absence also affected the engagement and energy of the other. 

\subsection{Self-Expression}

We saw interns express their feelings, values, and identities through many mediums and moments during the internship. Self-expression was particularly evident through the creative activities we facilitated as well as in reflection conversations. 

\subsubsection{Creative technology activities}

Through the creative technology activities, interns expressed aspirations and concerns about the future, personal accomplishments, journeys of healing from trauma, personal challenges they were trying to overcome, and feelings about loved ones. Their works touched upon themes of forgiveness, love, overthinking, resilience, happiness, freedom, connecting with family, and hopes for well-being.

We noticed that interns had unique, consistent affinities for particular forms of creative expression. One intern resonated more with verbal and movement-based self-expression, and the other found it easier to express themselves visually or with physical materials. Both interns were drawn to musical expression. These preferences impacted how the interns felt about activities that focused on a particular form of expression. For example, during a brainstorming activity prompting interns to think of and write down activity ideas (resulting in sticky notes, as seen in Figure \ref{fig:stickynotebrainstorm}), the latter intern expressed having trouble thinking of ideas. Even when facilitators suggested they try drawing or verbalizing instead of writing their ideas, they continued to feel discouraged. They later expressed that the experience made them question whether they were a good fit for the program overall. During a visual-focused brainstorm activity (Crazy 8s, as seen in Figure \ref{fig:crazy8brainstorm}), that intern appreciated the focus on visually expressing their ideas using a variety of materials and mediums of expression. The other intern, however, got frustrated by how hard it felt to express their thoughts without many words. 

The interns' self-expression preferences appeared to guide the digital activities they prototyped and designed later in the internship. The intern who preferred expressing themselves with visual and physical materials experimented with painting, building a miniature terrarium, and creating digital images, while the other intern focused on poetry and dance. This remained consistent through to their final designed activities (with the addition of music for both), with the former creating an activity through which youth use digital music tools to "express their feelings and calm down," followed by a calming activity of making art (aided by a generative AI-based tool) while listening to the music they created. The other intern created a guide for writing a poem about your feelings, setting it to music, and recording a dance to express what is being conveyed through the words and sounds. 

When exploring generative AI technology, we observed that tool design impacted the interns’ creative agency. We highlight one intern’s experience using MusicFX DJ \cite{MusicFXDJ} and Suno \cite{SunoAIMusic}. MusicFX DJ asks users to enter multiple text prompts (e.g. genres, instruments, or moods), and generatively blends these together into an instrumental track (e.g. Figure \ref{fig:music-activities}). The track plays continuously and the user can actively modify it while playing. Suno, by default, asks users for a single prompt input and generates complete songs, with layered instrumentals and generated lyrics. While somewhat hidden, each song has an edit button for additional iteration. When using MusicFX DJ, we observed the intern explore the tool with curiosity and intention. They chose "jazz," sharing out loud that they like to listen to it when they want to relax or calm down. Next, they added “drill beats” to see what would happen when this genre was combined with jazz. They felt this effect didn’t connect with the calming emotion they were trying to convey, so they adjusted a slider in the tool to lessen the intensity of the drill beats prompt on the output. When using Suno, they shared that they were “done” creating soon after entering their first prompt. We listened to the music together and asked questions about whether the generated lyrics connected with a personal experience. After reflecting on this, the intern wanted to change some of the lyrics, but didn't think this was possible within the interface. We pointed them to the “Edit” feature, but observed subsequent frustration in attempting to revise already complete lyrics that were not in their own words. This was also notable because of their affinity for non-verbal forms of expression. While both the tools offered technical features to support iteration, we found that the design of Suno’s interface and technical capabilities, particularly that it generated a complete song with minimal input from the user, became a barrier for the intern’s personal expression and exploration of the tool.

\subsubsection{Reflective conversations with facilitators} \label{reflective-conversations}

The interns frequently expressed how they were feeling to us, particularly when they were processing challenges they had experienced in their personal life or during the internship. 

Although morning check-ins generally started with an answer of "Fine" or "Good," they would go on to elaborate about the different things they were dealing with outside of work (such as interpersonal conflicts that had happened overnight, or things they were nervous about doing later). We observed that setting aside dedicated time and sitting in a separate space dedicated to sharing and reflection (Figure \ref{fig:couches}) created an environment in which the interns felt comfortable expressing how they were feeling. In the conversations, we focused on listening, validating their emotional responses and empathizing with their experiences, only occasionally offering guidance (drawing on one facilitator's experience as a CASA). We noticed that after having had this check-in, the interns were more energetic and present in activities. When challenges were particularly heavy and impacted their engagement, we used what we learned in the check-ins to adjust activities and provide emotional support throughout the day. 

Challenges with activities and tools introduced in the internship sometimes brought up hard emotions for the interns, particularly anger with themselves or the activity, embarrassment, or a feeling of overwhelm. In these heightened moments, we followed the intern's lead, offering optional space for them to reflect and process in the moment or later on. Usually, after taking some time to disengage (such as by going on a walk outside), the intern wanted to talk about how what had been coming up for them. They sometimes also chose to share in our group reflection time at the end of the day, which often led the other intern to share connected feelings and experiences of their own. This peer support seemed to help the interns feel seen, connected, and safer expressing their emotions in general. 

\subsection{Exploration}

The interns appeared to explore a range of potential interests and goals during the internship, frequently inspired by connective experiences they had with diverse people and fields. The activity design process also seemed to facilitate self-guided, deeper exploration, but came along with conceptualization challenges. 

\subsubsection{Connection-building experiences}

In exploration activities, the interns seemed to be particularly impacted by meeting people who had diverse interests and career journeys. In subsequent reflections, the interns would often share how this had sparked a new or rekindled interest that they may want to explore in the future, such as science, coding, drawing, making music, and combining multiple disciplines together. The value of the new connections they had formed came up in the final reflection activity, in which the interns shared that they "learned from so many different people, with so many backgrounds and focus areas" and messages such as "I am keeping myself on my journey. You can see what other people are doing and learn from them." They also emphasized the importance of "surrounding [themselves] with people who were on paths with similar goals," and wanting to "remember connections with people [they] met here."

The interns also appeared to engage in self-exploration in settings where they were connecting with topics rather than people (such as attending larger presentations and museum trips). For example, after a trip to a local science museum, one intern shared that they were now newly interested in taking engineering classes. The other intern shared that the trip to the museum reminded them of their childhood interest in space exploration. In the final internship activity, we reminded them of these conversations and they elaborated on some potential future interests. They added "space exploration," "robotics," "engineering," "bioengineering," and "trying different classes in college" to the list of goals and interests they were interested in exploring. 

\subsubsection{Open-ended activity design process}

The technology co-design process appeared to enable the interns to more freely explore, but also introduced abstraction that interns found difficult to conceptualize. 

As interns engaged in the creative activities in the first few weeks, they sometimes seemed to measure their success in comparison with the other intern or what they thought we wanted (as opposed to intrinsic goals). When prototyping their own technology-based activities, the interns appeared more comfortable working at their own pace and exploring mediums they were drawn to (see Figure \ref{fig:prototypes} for example prototypes). This additional autonomy seemed to reduce the pressure they put on themselves to achieve a particular outcome. During this prototyping stage, we observed the interns have moments of calm focused creation and express a sense of pride in their work. 

At the same time, while the interns were able to envision how other youth could use digital tools in new ways to express themselves and self-regulate, they had trouble breaking this down into steps for other youth to follow. We encouraged them to think about the 'recipe' or instructions they would give someone else, or to document their process of creating an example. However, both interns had trouble with this meta-cognitive step, and experienced moments of frustration when tried to provide encouragement and supplementary guidance. 

Despite their challenges with the activity design process, both interns eventually succeeded in envisioning new activities. Both interns talked about being motivated by designing something that other young people might actually use. Additionally, one intern shared that designing for other youth felt related to how teachers design activities for their students. In the final reflection activity, this intern expressed an interest in doing "something in education" such as becoming a "dance teacher" and shared that this interest was inspired by their experiences working on the design project. 

\subsection{Self-Efficacy}

As interns encountered a new environment, new technologies, and new forms of sharing and expressing themselves, they experienced challenges and moments of frustration. Reflection conversations enabled us to understand their internal experience and support them in building their belief in themselves and resolve to not give up in the face of obstacles.

\subsubsection{Challenging experiences with technology}

When engaging with technology to express aspects of their identities and experiences, interns sometimes struggled with working through challenges they were having with tools and activities. Challenges included adapting to new device and app interfaces that were confusing or did not respond as expected, interacting with interfaces and activities that required reading or writing (particularly for one intern who had dyslexia and ADHD), and engaging in reflection and iteration during the activity design process. These instances frequently brought up feelings of frustration for the interns, either with the activity or with themselves. We observed that when challenges with new technology arose, interns sometimes internalized them as issues with themselves as learners or technology users rather than faults in the design of the technology. In one instance, an intern was trying to use a generative AI image generation tool to create a personal vision for the future. Observing that the main figure looked very different from the intern (different skin tone, hair style, age), one facilitator asked whether the intern was happy with how the tool had depicted them or if there was anything they wished were different. The intern became frustrated, because, as we subsequently understood through a debrief conversation, they felt we were implying that the character did not look like them because they had used the tool incorrectly. 

\subsubsection{Affirming conversations with facilitators}

In reflection conversations after challenging experiences, interns frequently brought up internal goals they were already working towards related to perseverance, particularly around believing in themselves, combating perfectionism, and not giving up in the face of obstacles. With this context, we consciously tried to name, in real time, the effort we saw interns putting into activities that were challenging for them. We tried to highlight not only their successes but also the energy they put into not giving up when they felt frustrated or stuck. In our final reflection discussion, the interns emphasized that one of the main things that they learned during the internship was "I can do whatever I put my mind to!" When thinking about what they wanted to take away from the internship experience, they shared the personal reminder "Even though I get anxious, I can still do it! Just take a breath and come back." 

While the interns shared a bit about the specific technology and skills they had learned (e.g. "Using canva!", "Our presentations! We each delivered our messages in our own way." and "creating the activity"), when they were asked to reflect on the things they learned, accomplished, and wanted to remember from the internship, the interns primarily shared personal reminders of self-efficacy and resilience, such as "don't give up on yourself," "don't belittle yourself," and "unleash and unlock what I can do for myself." 

\section{Discussion}

Our results show how a technology-based internship designed to foster belonging for foster-involved youth can impact self-efficacy, self-expression, and exploration of personal interests and goals. Here, we contextualize our core findings within HCI literature on engaging youth in technology-based experiences. We then outline high-level recommendations for technologists and service providers aiming to create safe and welcoming opportunities for foster-involved youth to creatively express themselves with technology and engage in the co-design of new technological systems. 

\subsection{Core Takeaways \& Implications for HCI}

\subsubsection{Welcoming youth into technology-based experiences} HCI literature has underscored the importance of fostering inclusion in computing, but has often not outlined or examined the specific design factors that support young people in seeing themselves as members of a computing environment \cite{oguine_inclusion_2025}. In this study, we illustrate specific ways that programs can intentionally welcome and include minoritized youth as full participating members in unfamiliar computing environments. In particular, we noted the importance of ensuring that youth played an active role in shaping community norms, experienced unmediated and autonomous access to resources and the workspace (e.g. through building access, and accessible food), and had tangible manifestations of their inclusion in the community (e.g. through ID cards and dedicated personal workspaces, and opportunities to personalize their space). We found that these ingredients helped foster-involved youth show up and express themselves authentically, emotionally regulate, and engage in self-directed exploration of new environments and experiences. 




\subsubsection{Creating opportunities for youth to express themselves with technology in diverse and self-guided ways} Researchers have highlighted the importance of youth engaging in hands-on experiences with computing guided by their interests and identities \cite{resnickLifelongKindergartenCultivating2018, von_briesen_interventions_2025, codding_positionality_2019}, and discussed how these experiences can foster a sense of belonging in computing environments \cite{von_briesen_interventions_2025, codding_positionality_2019}. In our study, we show that the reverse can also occur---designing for belonging can enable  youth to engage in authentic self-expression and self-guided exploration with technology. Furthermore, we show how supporting youth in choosing how to express themselves can enable deeper engagement and allow them to explore their interests and goals. This extends literature on utilizing technology to support foster-involved youth with expressing themselves \cite{kumarConnectingComicsDesign2025, kumarCultivatingSupportiveSphere2025}, which has focused on providing pre-determined pathways for self-expression.  



\subsubsection{Supporting youth-led technology co-design} Existing HCI literature has solely engaged foster-involved youth in guided co-design experiences \cite{ezimoraReflectionsFosterYouth2025, kumarCultivatingSupportiveSphere2025, kumarConnectingComicsDesign2025}. In our co-design project, the open-ended design process enabled youth to more deeply explore their interests and career goals. At the same time, the project concept initially led to confusion and frustration, perhaps because the prompt to create an 'activity' felt ambiguous and abstract. Youth seemed to struggle with grasping their role as activity creators, maybe because they were used to being in spaces where someone else provided the instruction (e.g. school). However, through additional facilitation support from us and tenacity from the interns, they successfully designed novel activities that spoke to their unique perspectives on digital technology. Our results highlight that foster-involved youth who have had limited opportunities to freely explore technology and make their own decisions \cite{geenenTomorrowAnotherProblem2007, powersPerspectivesYouthFoster2018} can navigate open-ended and self-directed technology design projects with appropriate scaffolding to bolster their agency and autonomy. 



\subsubsection{Providing emotional support during technology experiences} Trauma-informed care frameworks advocate for the importance of fostering psychological safety and avoiding retraumatization to ensure youth with trauma experiences feel safe and are able to engage in activities \cite{fathallahTraumaResponsivenessDesign2024, salazarAuthenticallyEngagingYouth2021}. In order to support foster-involved youth in fully engaging in technology-based programs, we add that it is necessary to support them with processing the emotions that come along with learning and overcoming obstacles with technologies. In our program, we saw youth experience heightened emotions triggered both by difficulties in their personal lives as well as challenges they encountered with internship activities, particularly in trying to express themselves with digital tools. These challenges brought up feelings of frustration and connected to ongoing internal struggles interns had with being self-critical, doubting their abilities, and giving up. While we did not explicitly design the internship to help youth regulate and cope with negative emotions, we saw that this was a prerequisite to supporting youth with learning new digital tools and using them for self-expression and co-design. 



\subsubsection{Supporting critical reflections on technology} In computing education settings, fostering reflection and discussion on the potential benefits and harms of technology has been shown to increase sense of belonging for underrepresented students \cite{von_briesen_interventions_2025}. We add that without a critical reflective lens of technology, obstacles with technology can lead youth to question their capabilities as technologists and disengage from activities. During our program, youth faced challenges expressing themselves with digital technology, resulting in moments of significant frustration. When a digital tool yielded unexpected results, we observed youth sometimes misplacing blame on themselves. We hypothesize that these internalized frustrations may have been alleviated if we had better supported the interns in understanding technology as \textit{designed systems} which incorporate the biases and assumptions of designers and can lead to surprising or undesirable behavior.

\subsection{Design Recommendations}

We share the following design recommendations for future technologists and service providers aiming to create safe and welcoming opportunities for youth to express themselves and engage in co-design with technology, particularly for youth in foster care. 

\begin{enumerate}
    \item \textbf{Welcome youth into technology-based experiences with intention} to ensure youth feel safe, invited, and accepted in the environment and engage as their full selves. This includes giving youth clear manifestations of their inclusion, anticipating their needs, and supporting them in exercising autonomy and agency in the environment. 
    \item \textbf{Prepare to emotionally support youth as they encounter challenges}, particularly with new technologies. This includes ensuring that facilitators have the time, capacity, and training to provide a safe and affirming container for youth to express how they're feeling and support them in processing and coping in ways that preserve their autonomy. 
    \item \textbf{Engage youth in critically reflecting on technology as designed systems}, as a way to prevent youth from blaming themselves for unintended technology behavior. This includes building an understanding that new technologies might not work exactly the way we expect them to, and that tools are designed with some audiences in mind while leaving out others.
    \item \textbf{Provide youth with opportunities to engage in self-guided expression and exploration with technology}, to support them in building a deeper understanding of themselves and the world. This includes creating emergent opportunities for youth to engage with diverse digital tools and technologists and envision new ways technology could support youth like themselves. 
    \item\textbf{Couple open-ended technology co-design activities with appropriate scaffolding} to help youth navigate ambiguity and take on new leadership roles. Strategies can include providing clear and varied examples, supporting youth in determining how they want to navigate challenges, being ready and available if youth want support with problem-solving or working through frustrations, and adapting the project or activities based on youth needs.
\end{enumerate}

\subsection{Limitations and Future Work}
The takeaways of this paper may not be relevant to supporting all youth, or all foster-involved youth, given that this program was conducted with two youth in a particular geographic region. We hope that in the future, similar technology-based programs are run with other minoritized youth with diverse identities and from different regions to unpack the role of culture and identity in youths' experiences. Additionally, because this work was grounded in our reflections as program facilitators, it is inherently shaped by our belief systems and experiences. However, as a counterbalance, we drew heavily from verbatim intern reflections in the analysis process. Another limitation is that while we saw the interns build skills, beliefs, and connections during the internship, we did not track whether these still felt meaningful to them after the internship ended. More studies should explore the long-term impact of similar technology-based experiences. 

\begin{acks}
Acknowledgments will go here if the paper is accepted. 
\end{acks}

\bibliographystyle{ACM-Reference-Format}
\bibliography{references, manual_references}

\appendix
\section{Internship Job Description}\label{job-description}

\subsection{Intern's Daily Responsibilities}

Based on both interns’ schedules and needs we will determine the breakdown of in person and remote work, but we hope to have you working at [lab redacted for review] up to 20 hours a week. On a day to day basis, you will participate in a variety of creative and technology-related activities. You do not need to have any creative or technical skills before entering the program, we will make sure you have all the support you need to be successful.

Below is a rough overview of what we will do over the 5 week internship:
\begin{enumerate}
    \item Week 1: Exploring applications of creative technology for personal expression and creativity, emotion processing and regulation, and building community connections.
    \item Week 2: Personal reflection and identifying interest areas for your main project.
    \item Week 3 - 4: Imagining and experimenting on ideas for your main project.
    \item Week 5: Sharing and reflecting on what you’ve worked on.
\end{enumerate}

Over the 5 week internship, we hope to accomplish the following goals:
\begin{enumerate}
    \item You feel safe and comfortable expressing yourself.
    \item You learn strategies for design and problem solving.
    \item You feel a sense of control over the technologies that are around you in your life.
    \item You experience the joy of creating things that you care about.
\end{enumerate}

\subsection{Intern expectations}

Throughout the internship, we expect you to:
\begin{enumerate}
    \item Pay attention to how you're feeling. If you feel overwhelmed, stressed, or triggered by something happening in or outside of work, please take care of yourself first. If you are comfortable, we are here to listen and figure out how to get you the support you need, whether that means modifying assignments or helping you find resources outside of work.
    \item Attend all meetings and working sessions. That being said, we know life can be hard or surprising. If you have a conflict or are running late, just let us know and we'll help make sure you are brought up to speed on what you miss.
    \item Help foster an environment that is safe and welcoming, so that both you and the other intern feel comfortable sharing your perspectives and needs.
\end{enumerate}

\subsection{Dress code}

At the [lab redacted for review], people wear many different types of clothes – from button down shirts and slacks, to t-shirts and jeans, to dinosaur costumes. You should feel free to wear whatever makes you comfortable, as long as it does not contain hate speech/symbols or expose private parts of your body. We recognize that dress codes can perpetuate harmful stereotypes and contribute to discrimination against marginalized communities, so we hope to collaboratively define the expectations with you so that you and the other intern both feel comfortable and able to fully express yourselves.

\section{Expectations \& Community Norms}\label{community-norms}

\subsection{Intern expectations}
\begin{enumerate}
    \item Pay attention to how you're feeling. If you feel overwhelmed, stressed, or triggered by something happening in or outside of work, please take care of yourself first. If you are comfortable, we are here to listen and figure out how to get you the support you need, whether that means modifying assignments or helping you find resources outside of work.
    \item Attend all meetings and working sessions. That being said, we know life can be hard or surprising. If you have a conflict or are running late, just let us know and we'll help make sure you are brought up to speed on what you miss.
    \item Help foster an environment that is safe and welcoming, so that both you and the other intern feel comfortable sharing your perspectives and needs. 
\end{enumerate}

\subsection{Supervisor expectations}
\begin{enumerate}
    \item Check in before diving into work. Ask us how we feel at the start of each work day, and accept that how we feel comes into how we show up at work. 
    \item Be straightforward. Provide timely, constructive and honest feedback so we can grow. 
    \item Have drinks and snacks available. This helps meet our needs and makes the space more comfortable.
    \item Provide personal space. Give us our own desks so no one is crowded together.
\end{enumerate}

\subsection{Community norms}
\begin{enumerate}
    \item Practice thinking from other people’s perspectives (how they might be feeling, what they might be going through) before making a judgement about their actions.
    \item Be patient with ourselves and others. We make mistakes and work at different paces, and what’s important is that we continue to try and reflect.
    \item Let everyone share their opinion. Keep an open mind, and if someone is being quiet, consider asking what they’re thinking. 
\end{enumerate}

\section{Weekly Internship Schedule} \label{schedule}

Each week, we provided the interns with a schedule of planned activities. Below is an example, from the first week of the internship.

\begin{figure}[hbt!]
    \centering
    \includegraphics[width=.4\textwidth]{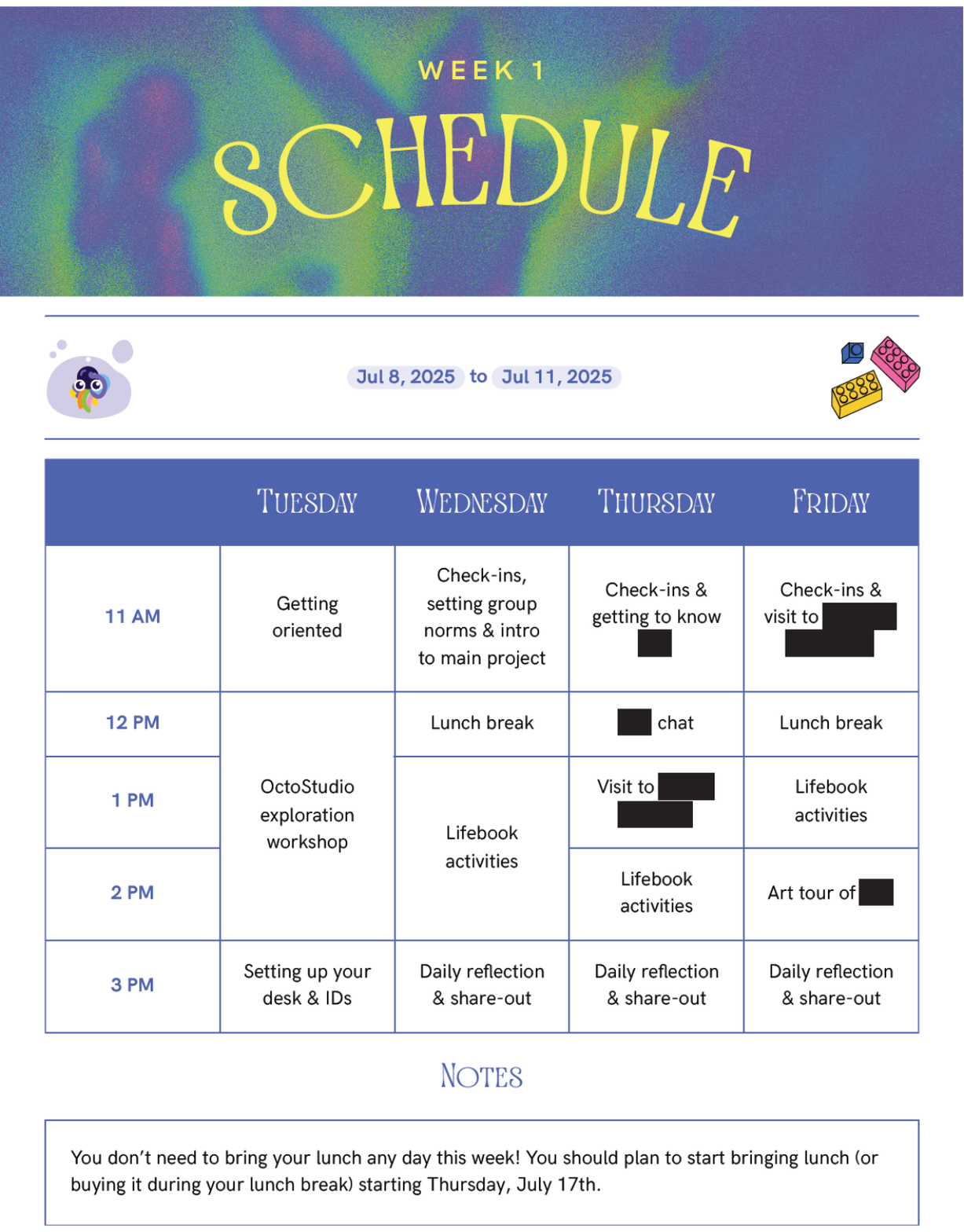}
    \caption{An example schedule, outlining the anticipated activities for the first week of the internship.}
    \Description{A PDF of a schedule for the first week of the internship, providing an hourly plan for each day of the week.}
\end{figure}

\end{document}